\documentclass[3p,preprint]{elsarticle}

\usepackage{hyperref}
\usepackage{bm}
\usepackage{placeins}
\usepackage{array}
\usepackage{caption}

\usepackage{framed} 
\usepackage{multicol} 
\usepackage{nomencl} 
\makenomenclature
\renewcommand*\nompreamble{\begin{multicols}{2}}
\renewcommand*\nompostamble{\end{multicols}}

\usepackage{color}
\usepackage{relsize}
\usepackage[utf8]{inputenc}

\journal{Engineering Fracture Mechanics}

\begin{document}

\begin{frontmatter}

\title{An efficient phase-field model for fatigue fracture in ductile materials}


\author[mymainaddress]{Martha Seiler}

\author[mymainaddress]{Thomas Linse}

\author[mysecondaryaddress]{Peter Hantschke}

\author[mymainaddress,mythirdaddress]{Markus K\"{a}stner\corref{mycorrespondingauthor}}
\cortext[mycorrespondingauthor]{Corresponding author}
\ead{markus.kaestner@tu-dresden.de}

\address[mymainaddress]{Chair of Computational and Experimental Solid Mechanics, TU Dresden, Dresden, Germany}
\address[mysecondaryaddress]{Structural Durability Group, TU Dresden, Dresden, Germany}
\address[mythirdaddress]{Dresden Center for Computational Materials Science (DCMS), TU Dresden, Dresden, Germany}

\begin{abstract}	
Fatigue fracture in ductile materials, e.\,g. metals, is caused by cyclic plasticity. Especially regarding the high numbers of load cycles, plastic material models resolving the full loading path are computationally very demanding. 
Herein, a model with particularly small computational effort is presented. It provides a macroscopic, phenomenological description of fatigue fracture by combining the phase-field method for brittle fracture with a classic durability concept.  A local lifetime variable is obtained, which degrades the fracture resistance progressively. By deriving the stress-strain path from cyclic material characteristics, only one increment per load cycle is needed at maximum.
The model allows to describe fatigue crack initiation, propagation and residual fracture and can reproduce Paris behaviour.
\end{abstract}


\end{frontmatter}

\section{Introduction}

Fatigue fracture is one of the most common causes of failure in structures, while still being insufficiently predictable. The fatigue life of a repeatedly loaded structure can be divided into three stages \cite{radaj_ermudungsfestigkeit_2007}, \cite{pineau_failure_2016}: During the crack initiation stage slip bands are formed. Later, micro-structurally small cracks merge to macro cracks. The macro cracks first grow stably (crack propagation) until the residual cross section is overloaded and the structure fails (residual fracture). The influence of microstructure is especially significant for high cycle fatigue, because in this setting most of the lifetime is spent in the micro-structurally small regime, while for low cycle fatigue, plasticity is the determining factor \cite{pineau_failure_2016}.

For many applications it is sufficient to estimate the number of load cycles until crack initiation $N_\mathrm{frac}$. Thereby crack initiation is often defined as the appearance of a technical crack of several millimetres. For this purpose statistical durability concepts based on W\"{o}hler experiments are employed. Component W\"{o}hler curves describe $N_\mathrm{frac}$ given a certain nominal stress amplitude. Beyond that, from strain-controlled experimental settings it is possible to derive the characteristic cyclic behaviour of the material itself. The strain W\"{o}hler curves are e.~g. utilised within the local strain approach (LSA) \cite{seeger_grundlagen_1996}, which ranks among the more advanced durability concepts and is able to consider the influence of plastic strains. 

For highly optimised structures, on the other hand, the crack propagation stage has to be considered as well during lifetime prediction in order to design the structures economically. Usually, the defect tolerant approach \cite{suresh_propagation_1984}, \cite{campbell_damage_1972} is adopted: Target quantity is the number of cycles it takes a crack to grow from a just visibly detectable range to failure.

Fracture mechanics provides tools to describe this crack propagation, such as the stress intensity factor $K$ \cite{irwin_analysis_1957}. The Paris-Erdogan law \cite{paris_critical_1963}  applies this concept to fatigue: It links the cyclic crack propagation rate $\mathrm{d}a/\mathrm{d}N$ with the amplitude stress intensity factor $\Delta K$ by the power law
\begin{equation} \label{eq:Paris}
\frac{\mathrm{d}a}{\mathrm{d}N} = C\Delta K^m.
\end{equation}
The Paris parameters $C$ and $m$ are meant to be material parameters and therefore independent of geometry and load. 
Noroozi et al. \cite{noroozi_two_2005}, \cite{noroozi_study_2007} and Mikheevskiy and Glinka \cite{mikheevskiy_elasticplastic_2009} combine fracture mechanics with the LSA, deriving a crack driving force for fatigue crack propagation. Instead of a sharp crack, which in an elastic setting produces a stress singularity, they consider plasticity at the crack tip, which is incorporated by equivalent residual stresses. 

Besides statistical and analytical methods, crack propagation can be simulated numerically. Most commonly, the finite element method is used for the spatial discretisation on different scales \cite{kuna_numerische_2010}. The numerical approaches can be distinguished in terms of the geometrical representation of the crack. Sharp crack models generally use conforming meshes, i.~e. the crack propagates along element edges or faces, respectively. The kinetics of crack growth are e.~g. controlled in terms of nodal forces or cohesive zone models. As a consequence of the conforming meshes, the approaches are well suited for known crack paths but require topological updates for arbitrary unknown directions of crack growth. These restrictions have been partly relaxed by the extended finite element method (XFEM) \cite{moes_finite_1999} which allows for cracks within finite elements in terms of an enriched displacement approximation. 

While the simulation of evolving cracks using a sharp crack topology proves difficult in particular in 3D settings, diffuse representations of the crack path may offer significant benefits from a computational point of view. Therefore, the phase-field method which is conceptually similar to gradient damage models is currently subject of intensive research. It is able to model both crack initiation and propagation, as well as arbitrary crack geometries in a straightforward manner using an additional field variable to represent the crack. For a description of the standard phase-field model for brittle fracture, which will also be used here, see Miehe et al. \cite{miehe_phase_2010}, \cite{miehe_thermodynamically_2010}. A review on a wider range of phase-field formulations for fracture can be found in Ambati et al. \cite{ambati_review_2015}. 

Recently, several propositions to extend the phase-field method to fatigue \cite{boldrini_non-isothermal_2016, caputo_damage_2015, amendola_thermomechanics_2014, carrara_novel_2018, mesgarnejad_phase-field_2018} have been published. Representatively for the range of different approaches, two models which are able to reproduce Paris behaviour shall be highlighted here: Carrara et al. \cite{carrara_novel_2018} introduce a phase-field model for fatigue fracture in \textit{brittle} materials. The basic idea of their approach is that due to repetitive loading the crack resistance decreases, allowing cracks to evolve even far below the static crack resistance. The fracture toughness is modified depending on a measure of locally accumulated elastic strain energy density. Mesgarnejad et al. \cite{mesgarnejad_phase-field_2018} follow a similar approach, but link the lowering of the fracture toughness also to the phase-field, localising the degradation to the vicinity of the crack tip. 

Although the authors use different approaches, they are not yet overcoming one key challenge inherent to fatigue crack initiation and propagation: The immense computational effort related to the high number of load cycles. This issue is addressed in the present paper. In particular, we introduce an efficient phase-field model of fatigue fracture in \textit{ductile} materials, such as metals. Analogously to \cite{carrara_novel_2018}, it is based on the reduction of the critical fracture energy, but uses a different local fatigue measure. In contrast to brittle materials, fatigue crack propagation in ductile materials is caused by cyclic plastic deformations. The straightforward way to treat the problem with an elasto-plastic material model is numerically expensive. Therefore, a different approach is chosen. With the help of the LSA, a local \textit{lifetime variable} is introduced, accounting for cumulative elasto-plastic deformations. Since the stress-strain path within a load cycle is derived from material curves from cyclic experiments, the explicit simulation of each load cycle would be redundant and can therefore be avoided, saving  computational costs. 
In other words, instead of introducing a ductile phase-field model, an elastic, brittle phase-field formulation which considers the elasto-plastic origins of fatigue is presented. As cracks are described on a macroscopic scale, microscopic effects are not resolved explicitly, but are represented statistically in the cyclic material characteristics. Conveniently, the utilised characteristics are derived from standardised experiments for which a large data base is already available.

This combination of the phase-field method with the LSA is considerably more flexible than the mere LSA, which itself is not applicable to crack propagation at all. The newly introduced model also incorporates static fracture with prior fatigue damage due to the general character of the phase-field approach.

This paper is organised as follows: The new model is introduced in Section~\ref{sec:model}, beginning with the standard phase-field formulation for brittle fracture, followed by its novel extension to fatigue. Furthermore, a scheme for numerical implementation for mean stress free constant amplitude loading is described. The method is applied in one- and two-dimensional examples in Section~\ref{sec:NumEx}. Finally, conclusions are drawn in Section~\ref{sec:Conc}.

\begin{table*}[!t]
	\begin{framed}
		\nomenclature{$\Pi$}{Total potential}
		\nomenclature{$\Pi_\ell$}{Total potential for regularised crack}
		\nomenclature{$\Omega$}{Domain}
		\nomenclature{$\Gamma$}{Crack surface}
		\nomenclature{$\psi^\mathrm{e}$}{Elastic energy density}
		\nomenclature{$\psi^\mathrm{e}_+$}{Tensile part of $\psi^\mathrm{e}$}
		\nomenclature{$\psi^\mathrm{e}_-$}{Compressive part of $\psi^\mathrm{e}$}
		\nomenclature{$g$}{Degradation function}
		\nomenclature{$\mathcal{H}$}{Crack driving force}
		\nomenclature{$\boldsymbol{n}$}{Normal}
		\nomenclature{$\tilde{\boldsymbol{t}}$}{Boundary traction}
		\nomenclature{$\boldsymbol{u}$}{Displacement}
		\nomenclature{$\tilde{\boldsymbol{u}}$}{Boundary displacement}
		\nomenclature{$t$}{Time}
		\nomenclature{$A$}{Area}
		\nomenclature{$\partial\Omega$}{Boundary of $\Omega$}
		\nomenclature{$\mathcal{G}_\mathrm{c}$}{Fracture toughness}
		\nomenclature{$\lambda$}{1st Lam\'{e} constant}
		\nomenclature{$\mu$}{2nd Lam\'{e} constant}
		\nomenclature{$\ell$}{Characteristic length}
		\nomenclature{$K'$}{Cyclic hardening coefficient} 
		\nomenclature{$n'$}{Cyclic hardening exponent}
		\nomenclature{$\sigma'_\mathrm{f}$}{Parameter of strain W\"{o}hler curve}
		\nomenclature{$\varepsilon'_\mathrm{f}$}{Parameter of strain W\"{o}hler curve}
		\nomenclature{$b$}{Parameter of strain W\"{o}hler curve}
		\nomenclature{$c$}{Parameter of strain W\"{o}hler curve}
		\nomenclature{$E$}{Young's modulus}
		\nomenclature{$d$}{Phase-field}
		\nomenclature{$D$}{Lifetime variable}
		\nomenclature{$\alpha_0$}{Fatigue degradation threshold}
		\nomenclature{$\xi$ }{Fatigue degradation exponent}
		\nomenclature{$\alpha$}{Fatigue degradation function}
		\nomenclature{$\boldsymbol{\sigma}$}{Stress tensor}
		\nomenclature{$\sigma_\mathrm{el}$}{Equivalent stress}
		\nomenclature{$\sigma$}{Revaluated equivalent stress}
		\nomenclature{$\sigma_\mathrm{m}$}{Cycle mean of $\sigma$}
		\nomenclature{$\sigma_\mathrm{a}$}{Cycle amplitude of $\sigma$}
		\nomenclature{$\boldsymbol{\varepsilon}$}{Strain tensor}
		\nomenclature{$\varepsilon_\mathrm{el}$}{Strain corresponding to $\sigma_\mathrm{el}$}
		\nomenclature{$\varepsilon$}{Strain corresponding to $\sigma$}
		\nomenclature{$\varepsilon_\mathrm{a}$}{Strain corresponding to $\sigma_\mathrm{a}$}
		\nomenclature{$\varepsilon_\mathrm{a,el}$}{Elastic strain part corresponding to $\sigma_\mathrm{a}$}
		\nomenclature{$\varepsilon_\mathrm{a,pl}$}{Plastic strain part corresponding to $\sigma_\mathrm{a}$}
		\nomenclature{$P_\mathrm{SWT}$}{Damage parameter}
		\nomenclature{$i$}{Increment number}
		\nomenclature{$N$}{Number of load cycles}
		\nomenclature{$N_\mathrm{frac}$}{Load cycles until failure}
		\nomenclature{$h_\mathrm{min}$}{Minimum element size}
		\nomenclature{$\tilde{u}_\mathrm{a}$}{Boundary displacement amplitude}
		\nomenclature{$\tilde{F}$}{Force boundary condition}
		\nomenclature{$\tilde{F}_\mathrm{a}$}{Force amplitude}
		\nomenclature{$C$}{Paris parameter}
		\nomenclature{$m$}{Paris parameter}
		\nomenclature{$\Delta K$}{Amplitude of stress intensity factor}
		\nomenclature{$a$}{Crack length}
		\nomenclature{$L$}{Length}
		\nomenclature{$s$}{Control variable for time}
		\nomenclature{$T$}{Thickness of CT specimen}
		\nomenclature{$W$}{Dimension of CT specimen}
		\nomenclature{$K$}{Stress intensity factor}
		\printnomenclature
	\end{framed}
\end{table*}

\clearpage

\section{Phase-field model for fatigue fracture}
\label{sec:model}

Since the newly developed method is based on the standard phase-field method for brittle fracture, the underlying formulation is introduced first. Then the extension of the model to fatigue is described, which makes use of a local lifetime variable. It is shown how this variable is determined using the LSA and how the model can be implemented. 

\subsection{Phase-field model for brittle fracture}

\begin{figure} [h]
	\centering
	\includegraphics[width=0.2\linewidth]{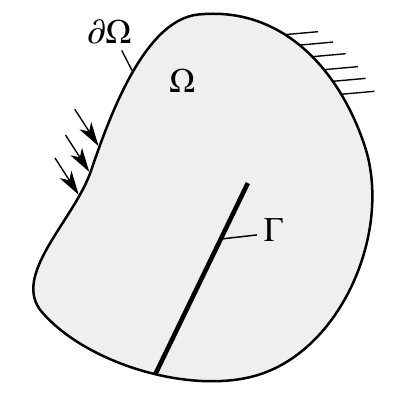}~a)
	\includegraphics[width=0.2\linewidth]{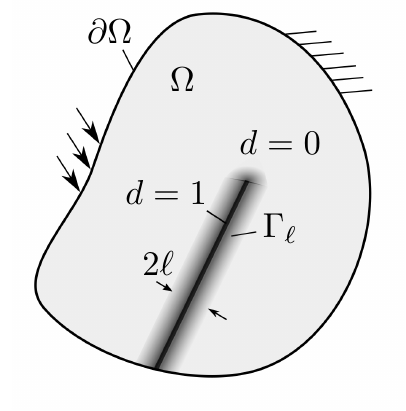}~b) 		 
	\caption{Fractured domain $\Omega$ with crack surface $\Gamma$. \textbf{(a)} Sharp representation of crack topology. \textbf{(b)} Regularized representation: The crack is described by the phase-field variable $d=1$, while $d=0$ represents undamaged material. The crack is regularized over the length scale $\ell$.
		\label{fig:Model_PF}}
\end{figure}

The phase-field method for brittle fracture is based on the Griffith-criterion \cite{griffith_phenomena_1921} of linear elastic fracture mechanics. It implies that a brittle crack can only propagate if the fracture energy, which is released during the formation of new crack surface, equals the critical energy release rate or fracture toughness $\mathcal{G}_\mathrm{c}$. A variational formulation for this criterion was proposed by Francfort and Marigo \cite{francfort_revisiting_1998}. Applying it by minimizing the total energy with regard to the displacement field and the crack geometry, arbitrary crack paths as well as crack initiation can me modelled without any further criteria.
The total potential $\Pi$ for a domain $\Omega$ with a fracture surface $\Gamma$, see Fig. \ref{fig:Model_PF}a, can be given by
\begin{equation}
\Pi = \int_{\Omega} \psi^\mathrm{e}( \boldsymbol{\varepsilon}) \, \mathrm{d} V + \int_{\Gamma} \mathcal{G}_\mathrm{c}\,\mathrm{d}A,
\end{equation}
excluding volume forces and boundary tractions for the sake of brevity. Assuming linear elasticity and small strains $\boldsymbol{\varepsilon}$, the elastic strain energy density can be written as 
\mbox{$\psi^\mathrm{e}=\frac{1}{2}\lambda\,\mathrm{tr}^2(\boldsymbol{\varepsilon})+\mu\,\mathrm{tr}(\boldsymbol{\varepsilon}^2)$}
with the elastic constants $\lambda$ and $\mu$. 

In order to enable a convenient numerical implementation, Bourdin et al. \cite{bourdin_numerical_2000} proposed a regularisation. To describe the crack topology, an additional field variable $d\in[0,1]$ is introduced, smoothly bridging the entirely intact ($d=0$) and totally broken ($d=1$) state. Kuhn and M\"{u}ller \cite{kuhn_continuum_2010} interpreted the states of the material as phases and applied the term phase-field. Approximating the sharp crack by a crack density $\gamma_\ell$ depending on a length scale parameter $\ell$, see Fig.~\ref{fig:Model_PF}b), the regularized energy functional can be written as
\begin{equation}
\Pi_\ell = \int_{\Omega} g(d)\,\psi^\mathrm{e}(\boldsymbol{\varepsilon}) \, \mathrm{d} V + \int_{\Omega} \mathcal{G}_\mathrm{c}\underbrace{\frac{1}{2\ell}(d^2+\ell^2|\nabla d |^2)}_{\gamma_\ell}\,\mathrm{d}V.
\end{equation}
Inspired by damage mechanics, a degradation function $g(d)=(1-d)^2$ is introduced which models the loss of stiffness due to the developing crack \cite{miehe_thermodynamically_2010}. Besides, it couples the mechanical field $\boldsymbol{u}$ and the phase-field $d$. The stress is given by
\begin{equation}
\boldsymbol{\sigma}=g(d)\frac{\partial\psi^\mathrm{e}}{\partial \boldsymbol{\varepsilon}}.
\end{equation}
From the variation $\delta\Pi_\ell=0$ one can derive the governing equations 
\begin{equation} \label{eq:goveq}
\vec{0}=\mathrm{div}\,\boldsymbol{\sigma} \qquad \qquad
d-\ell^2\Delta d = (1-d)\underbrace{\frac{2\ell}{\mathcal{G}_\mathrm{c}}\psi^\mathrm{e}(\boldsymbol{\varepsilon})}_\mathcal{H}
\end{equation}
subject to the boundary conditions $\boldsymbol{n}\cdot\boldsymbol{\sigma}=\tilde{\boldsymbol{t}}$, $\boldsymbol{u}=\tilde{\boldsymbol{u}}$ and $\boldsymbol{n}\cdot\nabla d=0$ with $\bar{\boldsymbol{t}}$ and $\bar{\boldsymbol{u}}$ being the prescribed boundary tractions and displacements.
In Eq.~(\ref{eq:goveq}), the crack driving force $\mathcal{H}$, which controls the evolution of the phase-field, can be identified.
Miehe et al.~\cite{miehe_phase_2010} interpreted the phase-field variable in the sense of damage and therefore introduced the crack driving force as the maximum energy density in time history
\begin{equation} \label{eq:irr}
\mathcal{H} = \frac{2\ell}{\mathcal{G}_\mathrm{c}} \max_{s\in [0;t]} \psi^\mathrm{e}(\boldsymbol{\varepsilon},s),
\end{equation}
ensuring local irreversibility for the phase-field. The effects thereof are discussed in \cite{linse_convergence_2017}. 
Physically motivated, the degradation can also solely be applied to the tensile part of the energy density 
\begin{equation}
\psi^\mathrm{e}_l(\boldsymbol{\varepsilon},d)=g(d)\psi^\mathrm{e}_+(\boldsymbol{\varepsilon})+\psi^\mathrm{e}_-(\boldsymbol{\varepsilon}),
\end{equation}
while the compressive range remains unaffected. The strain is thereby split into a volumetric and a deviatoric part \cite{amor_regularized_2009}, according to its principal components \cite{miehe_thermodynamically_2010} or the crack orientation \cite{steinke_phase-field_2018}.

\subsection{Extension to fatigue}

The following section addresses the question how fatigue can be modelled within the phase-field framework. Analogously to Carrara et al.~\cite{carrara_novel_2018}, the fracture toughness $\mathcal{G}_\mathrm{c}$ is reduced when the material degradation due to repetitive stressing precedes. Here, this process is described by a local \textit{lifetime variable} $D$. An additional scalar \textit{fatigue degradation function} \mbox{$\alpha(D)\in[\alpha_0,1]$} with $0<\alpha_0<1$ is introduced, which lowers the fracture toughness $\mathcal{G}_\mathrm{c}$ locally. The energy functional then reads
\begin{equation}\label{eq:Pi_l}
\Pi_\ell = \int_{\Omega} g(d)\,\psi^\mathrm{e}(\boldsymbol{\varepsilon}) \, \mathrm{d} V + \int_{\Omega} \alpha(D) \,\mathcal{G}_\mathrm{c}\frac{1}{2\ell}(d^2+\ell^2|\nabla d|^2)\,\mathrm{d}V. 
\end{equation}
The reduction of the total energy due to $\alpha(D)$ is meant to model the dissipation due to local cyclic plasticity. Demanding again $\delta\Pi_\ell=0$, the evolution equation of the phase-field under consideration of the irreversibility condition (\ref{eq:irr}) extends to
\begin{equation}\label{eq:ev_PF}
\alpha(D)\,d-\nabla\alpha(D)\cdot\ell^2\nabla d-\alpha(D)\,\ell^2\Delta d=(1-d)\max_{s\in [0;t]}\psi^\mathrm{e}({\boldsymbol{\varepsilon}})\frac{2\ell}{\mathcal{G}_\mathrm{c}}.
\end{equation}
The lifetime variable $D\in[0,1]$ is a history variable that is accumulated strictly locally. It can be interpreted as a damage variable with a special linear character: For $D=0$ a material point has experienced no fatigue loads at all, while $D=1$ means it has undergone all load cycles it can possibly bear before loosing its integrity, linearly spanning the lifetime in between. Its computation according to the LSA is described in Section~\ref{sec:calcD}. 

\begin{figure} [h]
	\centering
	\begin{tabular}{cc}
		\includegraphics[width=0.37\linewidth,trim={0.5cm 0.5cm 1cm 0.5cm},clip]{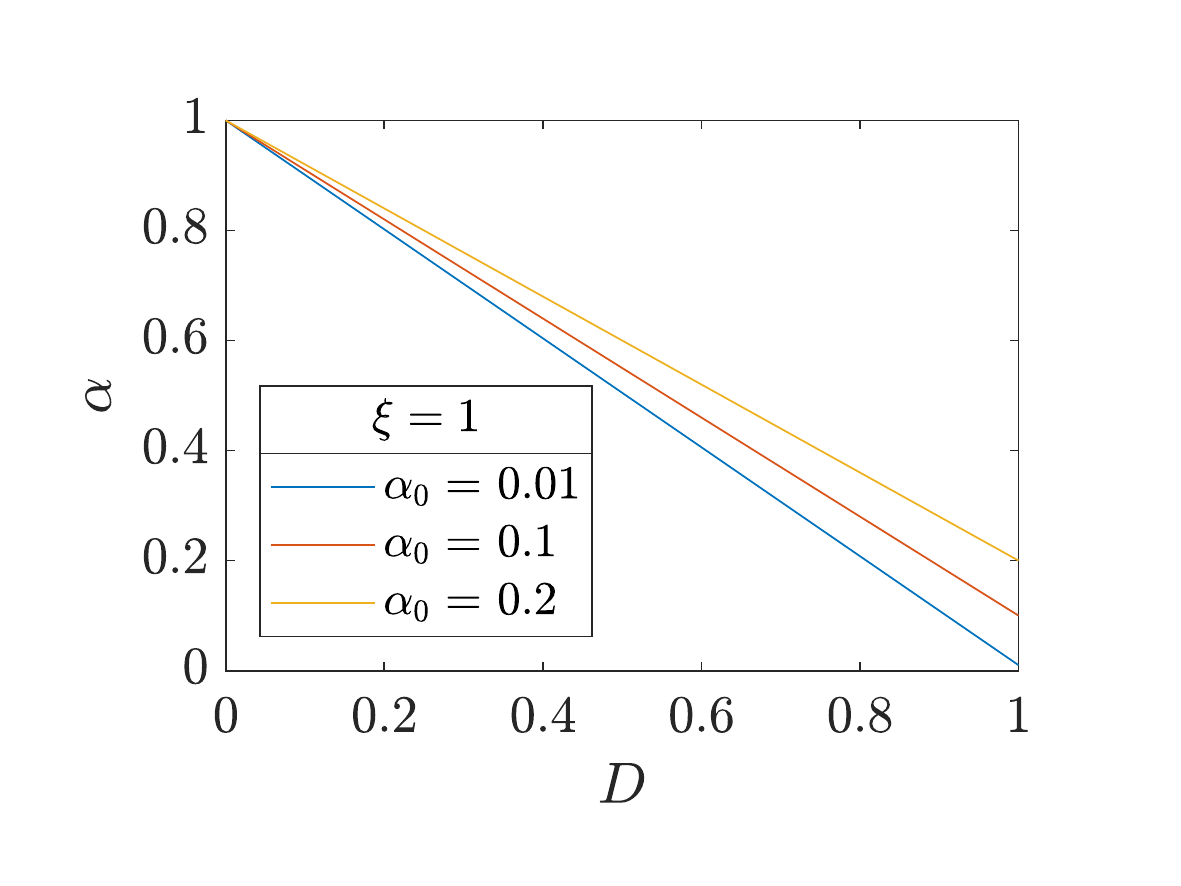} & 
		\includegraphics[width=0.37\linewidth,trim={0.5cm 0.5cm 1cm 0.5cm},clip]{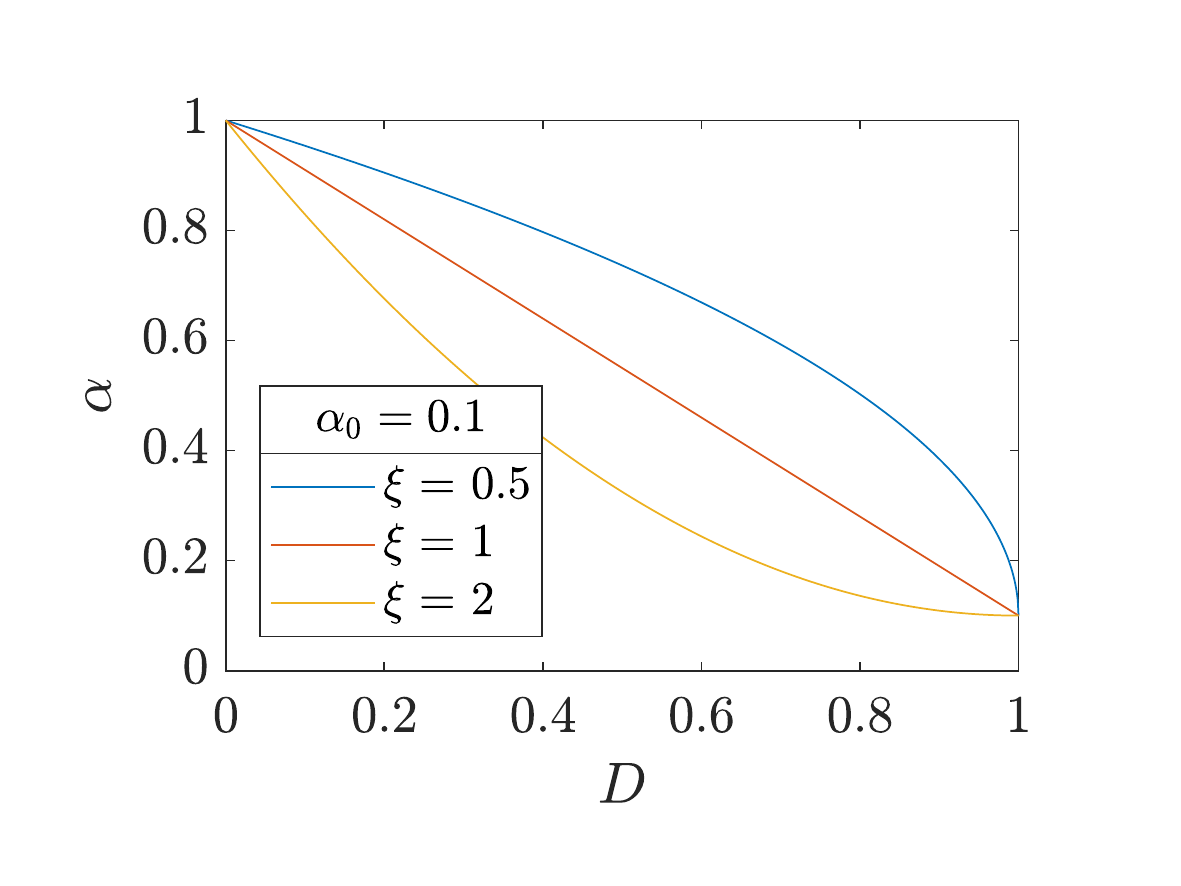} 
	\end{tabular}			 
	\caption{Fatigue degradation function $\alpha(D)=(1-\alpha_0)(1-D)^\xi+\alpha_0$, which degrades the fracture toughness depending on the local lifetime variable $D$. 
		\label{fig:Model_alpha}}
\end{figure}

On this basis, the fatigue degradation function
\begin{equation}
\alpha(D)=(1-\alpha_0)(1-D)^\xi+\alpha_0
\end{equation}
with the parameters $\alpha_0$ and $\xi$ is proposed, see Fig.~\ref{fig:Model_alpha} for an illustration. For $D=0$, the material has experienced no cyclic loads at all and therefore must have full fracture toughness, consequently $\alpha(0)=1$ must hold. 
The parameter $\xi$ controls the relation between $D$ and $d$. The threshold $\alpha_0$, on the other hand, offers a link to experiments on residual fracture in which the remaining fracture toughness of fatigued components is measured. It has to be larger than zero, since even cyclically damaged material at the end of its lifetime offers a certain resistance against crack propagation. 
The influence of both parameters $\alpha_0$ and $\xi$ on crack initiation and propagation is studied in Section~\ref{sec:NumEx}.

Since the phase-field variable is here interpreted as physical damage instead of as a marker for a potential crack, the irreversibility condition (\ref{eq:irr}) is adopted. However, a tension-compression split is not necessary, as will be explained further in Section~\ref{sec:calcD}.

\subsection{Local Strain Approach (LSA)}
\label{sec:calcD}

The computation of the lifetime variable $D$ follows the LSA \cite{seeger_grundlagen_1996},  sometimes also called notch strain concept, which is generally used for service life prediction of structures. Therein, experimental data can be applied to arbitrary specimen because of the assumption that same strains generate same damage in same material. In contrast to other fatigue concepts, the LSA is a material concept as it considers the local stress-strain path instead of a nominal stress quantity. It is therefore especially suitable for application in combination with the finite element method. 
In the classical sense, the LSA is evaluated at the point of the structure with the highest stress which is assumed to be decisive for component failure. Here, it is applied to each material point instead. 

Due to the local formulation of the LSA crack propagation can be described as crack initiation at a number of material points. Interestingly, this approach is similar to the crack propagation model of Noroozi et al. \cite{noroozi_two_2005}, who describe crack growth by applying the LSA blockwise to elements, interpreting crack growth as successive crack initiation.

\begin{figure} [p]
	\begin{minipage}[b]{0.52\textwidth}
		\includegraphics[width=\linewidth]{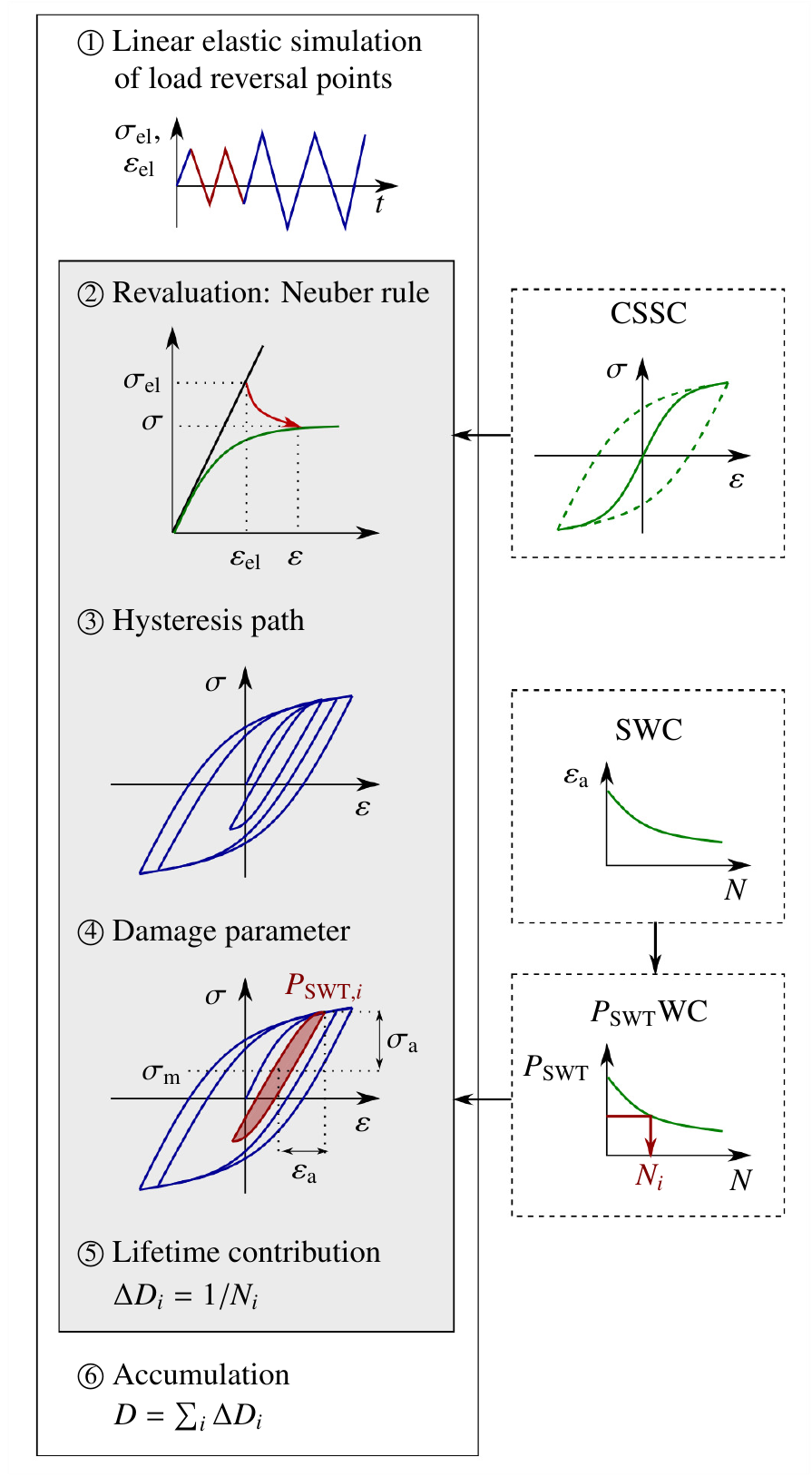} 
	\end{minipage}
	\hspace{0.5cm}
	\begin{minipage}[b]{0.48\textwidth}
		\includegraphics[width=\linewidth]{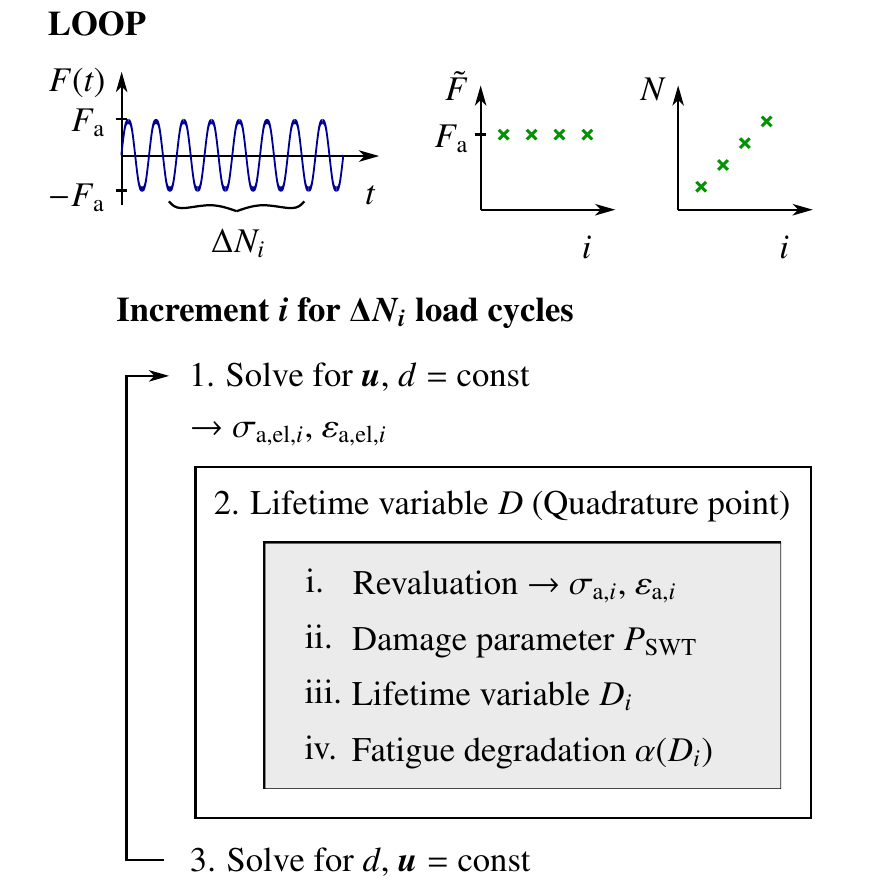} 		 
	\end{minipage}
	\begin{minipage}[t]{0.52\textwidth}
		\caption{Scheme for determination of the local lifetime variable $D$ using the local strain approach (shaded). The elastically determined stress-strain state is revaluated to the cyclic stress-strain curve (CSSC) using the Neuber rule. For each load cycle, the CSSC and amplitude values of stress and strain $\sigma_\mathrm{a}$ and $\varepsilon_\mathrm{a}$ define the stress-strain hysteresis, from which the damage parameter $P_\mathrm{SWT}$ can be determined. The theoretically bearable number of load cycles to fracture $N$ is then drawn from $P_\mathrm{SWT}$-W\"{o}hler curves ($P_\mathrm{SWT}$WC), which are derived from strain W\"{o}hler curves (SWC). It determines the lifetime contribution $\Delta D$ which can be accumulated linearly over all load cycles.
			\label{fig:Model_NSC}}
	\end{minipage}
	\hspace{0.5cm}
	\begin{minipage}[t]{0.48\textwidth}		 
		\caption{Scheme for numerical implementation. In a staggered algorithm, the fields $\bm{u}$ and $d$ are solved separately. In between, at each quadrature point, the local strain approach (shaded) is applied to the elastically simulated stress amplitude $\sigma_{\mathrm{a,el}}$. For constant load amplitudes, the boundary condition is kept to the amplitude value for all increments. Each increment covers $\Delta N$ load cycles.
		\label{fig:Model_SimScheme}}
	\end{minipage}
\end{figure}

The lifetime variable $D$ is computed load cycle wise, each load cycle $i$ contributing $\Delta D_i$. The computation scheme, adapted to application within the introduced model, is illustrated in Fig.~\ref{fig:Model_NSC}. 
It is based on the loading path. For each reversal point, marking a minimum or a maximum load of a cycle, a linear elastic simulation \textcircled{\smaller 1} determines the elastic stress-strain state. The LSA generally uses only scalar stresses and strains, as it is based on experiments in which only axial quantities are considered. It is common practice to use the von Mises equivalent stress in case of multiaxial stress states \cite{neuber_theory_1961}. 
The elastically determined equivalent stress $\sigma_\mathrm{el}$ as well as all stress and strain quantities derived from it are denoted by non-bold symbols.

\begin{figure} [h]
	\begin{minipage}[c]{0.5\textwidth}
		\includegraphics[width=\linewidth]{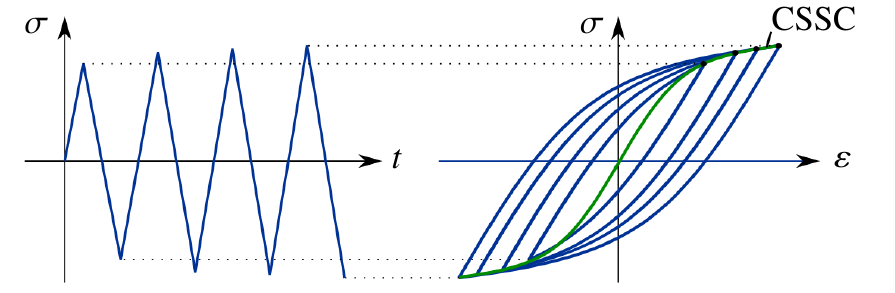}
	\end{minipage}
	\hspace{0.5cm}
	\begin{minipage}[c]{0.45\textwidth}
		\includegraphics[width=\linewidth]{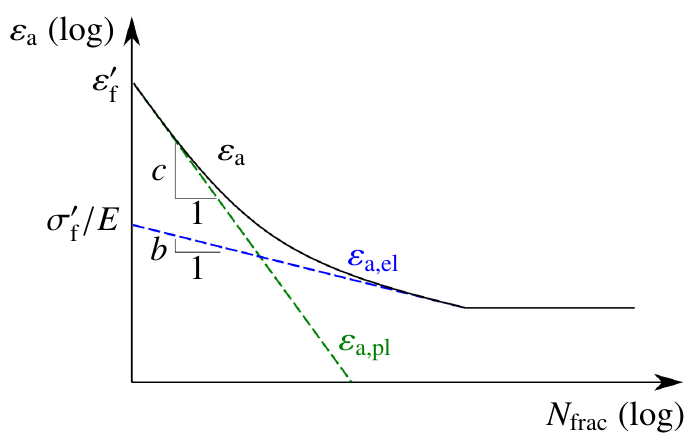}  		 
	\end{minipage}
	\begin{minipage}[t]{0.5\textwidth}
		\caption{ The cyclic stress-strain curve (CSSC) tracks the reversal points of a cyclic experiment and thereby the maxima and minima of the stress-strain hystereses. 
			\label{fig:Model_CSSC}}
	\end{minipage}
	\hspace{0.5cm}
	\begin{minipage}[t]{0.45\textwidth}		 
		\caption{Strain W\"{o}hler curve (SWC). Material characteristic, that displays the relation between the strain amplitude $\varepsilon_\mathrm{a}$ and the number of load cycles until failure $N_\mathrm{frac}$. Due to the strain-control of the experiment, the effects of elastic and plastic part $\varepsilon_\mathrm{a,el}$ and $\varepsilon_\mathrm{a,pl}$ can be considered.
			\label{fig:Model_DWL}}
	\end{minipage}
\end{figure}

First of all, stresses and strains are revaluated to elasto-plastic values \textcircled{\smaller 2} using the Neuber rule as a revaluation hypothesis and the cyclic stress-strain curve (CSSC) of the material. This curve is a material characteristic which is determined experimentally. As shown in Fig.~\ref{fig:Model_CSSC}, the CSSC describes stress-strain relationship for the reversal points under cyclic loading. Usually, it is recorded during an Incremental Step Test \cite{landgraf_determination_1969} after the material has reached cyclic stabilisation.
It is approximated by the Ramberg-Osgood model \cite{ramberg_description_1943}
\begin{equation}
\label{eq:RamOs}
\varepsilon_\mathrm{a} = \frac{\sigma}{E} + \left( \frac{\sigma}{K'} \right)^{1/n'}
\end{equation}
with the cyclic hardening coefficient $K'$ and the cyclic hardening exponent $n'$. In the phase-field framework, the the stress $\sigma$ is thereby the degraded stress $\sigma =g(d)\, \partial\psi^\mathrm{e}/\partial\varepsilon$.

According to the Neuber rule \cite{neuber_theory_1961}, the product of stress and strain has to be the same before and after the revaluation
\begin{equation}
\label{eq:Neuber}
\sigma_\mathrm{el}\varepsilon_\mathrm{el} = \sigma\varepsilon. 
\end{equation}
Using the Ramberg-Osgood ansatz (\ref{eq:RamOs}) 
\begin{equation}
\label{eq:RamOs2}
\sigma_\mathrm{el} \frac{\sigma_\mathrm{el}}{E} = \sigma \left(\frac{\sigma}{E} + \left(\frac{\sigma}{K'}\right)^{1/n'}\right),
\end{equation}
the revaluated stress $\sigma$ can be derived by solving Eq.~(\ref{eq:RamOs2}) numerically.
This revaluation is not equivalent to a plastic material model, it rather incorporates plasticity as the cause of fatigue crack growth in an empirical way. 

The hysteresis curves \textcircled{\smaller 3} between the reversal points are described by the Masing curve \cite{masing_eigenspannungen_1926} which corresponds to the CSSC scaled by factor 2.
To characterise the damaging effect of one full hysteresis $i$, the damage parameter by Smith, Watson and Topper \cite{smith_stress-strain_1970} \textcircled{\smaller 4} 
\begin{equation}
\label{eq:PSWT}
P_{\mathrm{SWT},i} = \sqrt{(\sigma_{\mathrm{a},i}+\sigma_{\mathrm{m},i})\varepsilon_{\mathrm{a},i}E}
\end{equation}
is computed. It is associated with the area inside one hysteresis loop and is therefore connected to the occurring dissipation.  $P_\mathrm{SWT}$ only depends on the stress and strain amplitudes $\sigma_\mathrm{a}$ and $\varepsilon_\mathrm{a}$ and the mean stress $\sigma_\mathrm{m}$, which shall not be considered here yet. One simulation with amplitude load is therefore enough to obtain $\sigma_\mathrm{a}$ and $\varepsilon_\mathrm{a}$ for each quadrature point, describing the damaging effect of the full hysteresis without a simulation of the entire loading and unloading path.

The material specific sensitivity to fatigue is included by strain W\"{o}hler curves (SWC). See Fig.~\ref{fig:Model_DWL} for a schematic illustration. 
The relation between the strain amplitude $\varepsilon_\mathrm{a}$ and the number of load cycles until fracture $N_\mathrm{frac}$ is given by
\begin{equation}
\label{eq:ManMor}
\varepsilon_\mathrm{a} = \varepsilon_\mathrm{a,el} + \varepsilon_\mathrm{a,pl} = \frac{\sigma'_\mathrm{f}}{E} (2N_\mathrm{frac})^b + \varepsilon'_\mathrm{f}(2N_\mathrm{frac})^c
\end{equation}
with parameters $\sigma'_\mathrm{f}$, $\varepsilon'_\mathrm{f}$, $b$ and $c$ according to Manson, Coffin and Morrow \cite{manson_fatigue:_1965}, \cite{coffin_study_1954}, \cite{lazan_cyclic_1965}. Due to the strain-controlled setting of the underlying material experiments, both the contributions of elastic and plastic strains can be considered.
By reformulating Eq.~(\ref{eq:ManMor}) for the damage parameter $P_\mathrm{SWT}$
\begin{equation}
{P_{\mathrm{SWT},i}}^2 = {\sigma'_\mathrm{f}}^2(2N_i)^{2b} + \sigma'_\mathrm{f} \varepsilon'_\mathrm{f}E(2N_i)^{b+c},
\end{equation} 
the damage parameter W\"{o}hler curve ($P_\mathrm{SWT}$WC) is derived. After equalling it to Eq.~(\ref{eq:PSWT}), the equation can be solved for $N_{i}$, which describes the number of load cycles the material could bear assuming continuous stressing with only the damage parameter $P_{\mathrm{SWT},i}$. The lifetime contribution $\Delta D_i$ \textcircled{\smaller 5} of the single load cycle then is
\begin{equation}
\Delta D_i = 1/N_i.
\end{equation}
According to the linear damage accumulation hypothesis, i.\,e. Miner rule \cite{miner_cumulative_1945},\cite{palmgren_lebensdauer_1924}, the lifetime contribution of all passed load cycles \textcircled{\smaller 6} can be added up to
\begin{equation}
D=\sum_i \Delta D_i.
\end{equation}
Hence, e.\,g. $D=0.5$ implies that half of the bearable load cycles have been experienced, given constant stress amplitudes. It becomes clear that $D$, describing expired life of a material point as a strictly local history variable, is of a very different character than the phase-field variable $d$, which represents the evolution and the topology of the actual physical crack driven by $D$.
This coupling is implemented in the phase-field model by the fatigue degradation function $\alpha(D)$. 

Within the model no split is performed on the energy density. Due to the symmetry of the CSSC even material points which experience compressive stresses for $\sigma=\sigma_\mathrm{a}$ (upper reversal point), would experience tensile stresses for $\sigma=-\sigma_\mathrm{a}$ (lower reversal point) and should therefore contribute to crack propagation.

In conclusion, applying the LSA for recording the fatiguing effect of the load history on the material brings along two crucial benefits in the matter of computational effort:  Firstly, due to the revaluation an elastic material model is applied instead of an elasto-plastic one. And secondly, the explicit loading path has not to be resolved. Instead, only the amplitude load of the cycle is applied. 

\subsection{Numerical implementation}

In the following, a scheme for implementing the model in a standard finite element analysis programme for phase-field simulations is proposed. See Fig.~\ref{fig:Model_SimScheme} for an illustration.
It follows the staggered approach by Hofacker and Miehe \cite{hofacker_continuum_2012} which decouples displacement $\boldsymbol{u}$ and phase-field $d$. In an iterative scheme, both are solved separately while freezing the other one, respectively. The computation of the lifetime variable $D$ can be inserted between the two, resulting in a three-step scheme. The damaging effect of the loading, described by $D$, is computed from the stresses and strains derived from the mechanical field. $D$ reduces the fracture toughness, which can again lead to crack evolution in the phase-field.

Starting simply with a steady oscillating load with constant amplitude and no mean load, the following implementation scheme can be applied. Instead of explicitly simulating the loading path with loading and unloading stage, the boundary condition is kept to the amplitude value, e.~g. $\tilde{F}=\tilde{F}_\mathrm{a}$ for a force-controlled experiment. To begin with, each increment $i$ is associated with one load cycle. Within the increment, an iteration over the fields takes place. At first, the problem is solved for the mechanical field, yielding the local stress and strain amplitudes $\sigma_{\mathrm{a,el},i}$ and $\varepsilon_{\mathrm{a,el},i}$. At each quadrature point independently, those are revaluated to the pseudo elasto-plastic quantities $\sigma_{\mathrm{a},i}$ and $\varepsilon_{\mathrm{a},i}$, which are then used to compute the damage parameter $P_{\mathrm{SWT},i}$ and the lifetime contribution $D_i$. $D_i$ yields the local fatigue degradation factor $\alpha_i$. Considering the updated distribution of the fracture toughness $\alpha\,\mathcal{G}_\mathrm{c}$ accounting for the damage due to the load cycle $i$, the problem is solved for the phase-field. This iteration is stopped by relative and absolute convergence criteria for both fields. 

For small to moderate crack propagation rates it is even possible to cover several load cycles in one increment: Because of the linear character of $D$, the lifetime contribution of $\Delta N$ load cycles is $\Delta D = \Delta N/N_i$. Here, a control for the simulated load cycles per increment, depending on the rate of change in the fracture energy, is applied in order to cover the slow crack evolution at the beginning as well as the abrupt development later in fatigue life properly. Alternatively, $\Delta N$ can be hinged on the number of staggered loops until convergence.

Since the LSA is designed for arbitrary loads with varying amplitudes, the proposed model can be generalised to those loading cases. This is out of scope of this publication though. See \cite{kuhne_fatigue_2018} for a description of the LSA for general loads and transient cyclic stress-strain behaviour.
Due to the general character of the phase-field formulation, static loading is included as a special case $\Delta N=0$. For the limit case of static load on previously unloaded material, the model exactly matches the standard static phase-field model for brittle fracture. Previous cyclic loads before the static loading stage, on the other hand, are also covered by the reduced fracture toughness.

\newpage 

\section{Numerical examples}
\label{sec:NumEx}
In the following, the validity of the method and its influencing factors are to be studied using various examples. For the sake of simplicity the method is applied to a one-dimensional bar first, demonstrating crack initiation. In order to study crack propagation, single-edge notched tests as well as a compact tension test are presented subsequently. The examples solely serve the purpose of qualitative verification and to demonstrate fundamental principles, a comparison with experimental results is out of scope of this publication.

\subsection{One-dimensional example}
\label{sec:NumEx_1D}

\begin{figure}
	\begin{minipage}{0.55\textwidth}
		\captionsetup{type=table}
		\caption{Parameters of the cyclic stress-strain curve (CSSC) according to the Ramberg-Osgood model \ref{eq:RamOs} and the strain W\"{o}hler curve (SWC) according to the  Manson, Coffin and Morrow approach \ref{eq:ManMor}, both for 42CrMo4 steel \cite{boller_vergleich_nodate}.		
			\label{tab:parameter}\newline}
		\centering
		\begin{tabular}{ll|llll}
			CSSC && SWC &&& \\
			\hline
			$K'$ & $n'$ & $\sigma'_\mathrm{f}$ & $\varepsilon'_\mathrm{f}$ & $b$ & $c$ \\
			2115 MPa & 0.195 & 1554 MPa & 1.447 & -0.086 & -0.710
		\end{tabular}
	\end{minipage}
	\hspace{0.5cm}
	\begin{minipage}{0.4\textwidth}
		\centering
		\includegraphics[width=\linewidth]{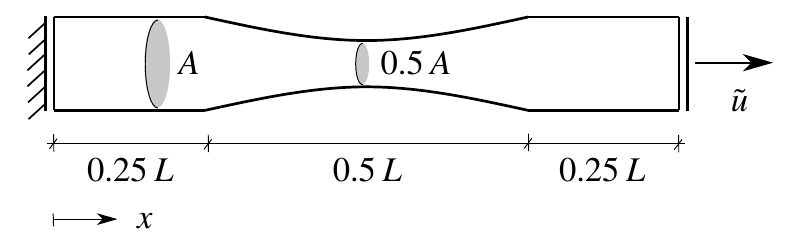} 		 
		\caption{One-dimensional setup: Bar with a quadratically reduced cross section in the middle. A constant displacement amplitude $\tilde{u}$ is applied.
			\label{fig:NumEx_geom}}
	\end{minipage}
\end{figure}

A one-dimensional bar of the length $L=100$~mm is considered. As shown in Fig.~\ref{fig:NumEx_geom}, its cross section is reduced in the middle from $A=A_0=100$~mm$^2$ to $0.5A_0$. The Young's modulus and the initial fracture toughness are set to $E=210$~GPa and $\mathcal{G}_\mathrm{c}=2.7$~N/mm, the characteristic length to $\ell= 1$~mm. The quadratic elements are of the size $h_\mathrm{min}=0.13\,\mathrm{mm}$ within the critical region. Unless otherwise stated, the parameters of the fatigue degradation function are chosen to $\xi=1$ and $\alpha_0=1\cdot10^{-8}\approx0$, whereat $\alpha_0$ is set to a small value instead of 0 to avoid numerical difficulties. The bar is virtually loaded with a displacement oscillation with an amplitude $\tilde{u}_\mathrm{a}$ and no mean load, which entails a constant displacement boundary condition $\tilde{u}=\tilde{u}_\mathrm{a}=0.13$~mm for the simulation. Due to an application project, a heat-treated steel, namely 42CrMo4, is used as an example here. The corresponding cyclic parameters are displayed in Tab.~\ref{tab:parameter}. 

\begin{figure} [p]
	\centering	
	\setlength{\tabcolsep}{0pt}
	\begin{tabular}{clcl}
		\includegraphics[width=0.48\linewidth,trim={0.5cm 0.5cm 1cm 0.5cm},clip] {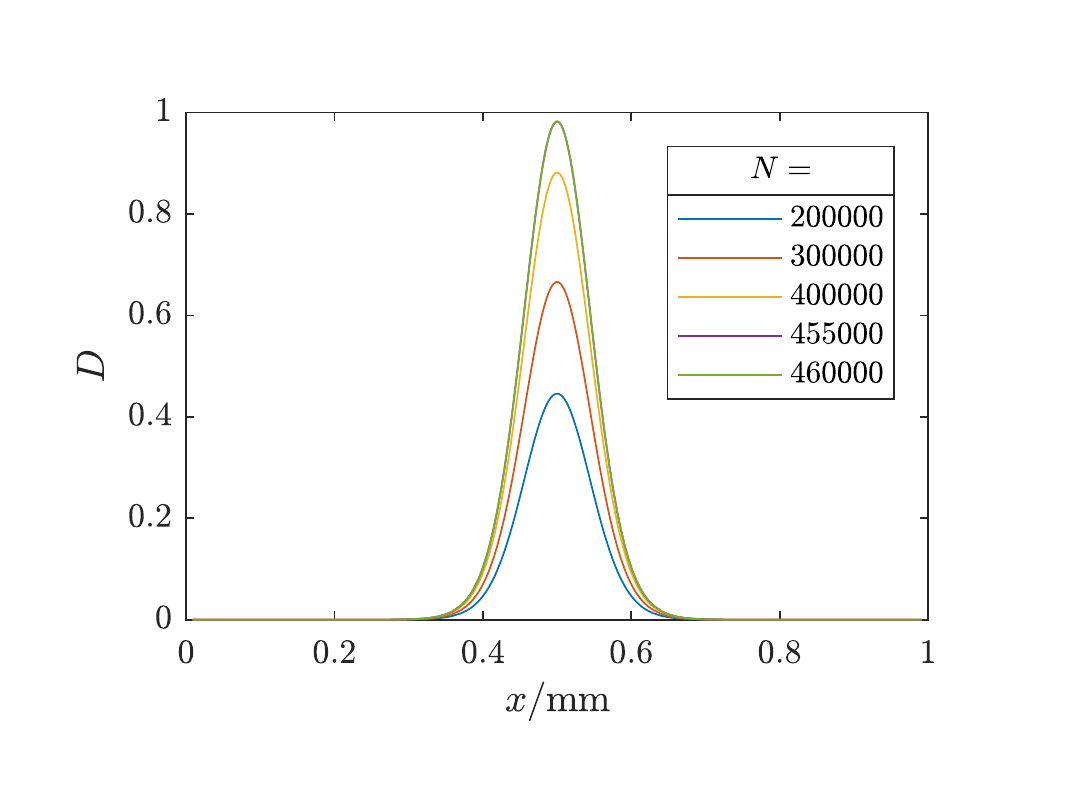} 	& a) &
		\includegraphics[width=0.48\linewidth,trim={0.5cm 0.5cm 1cm 0.5cm},clip] {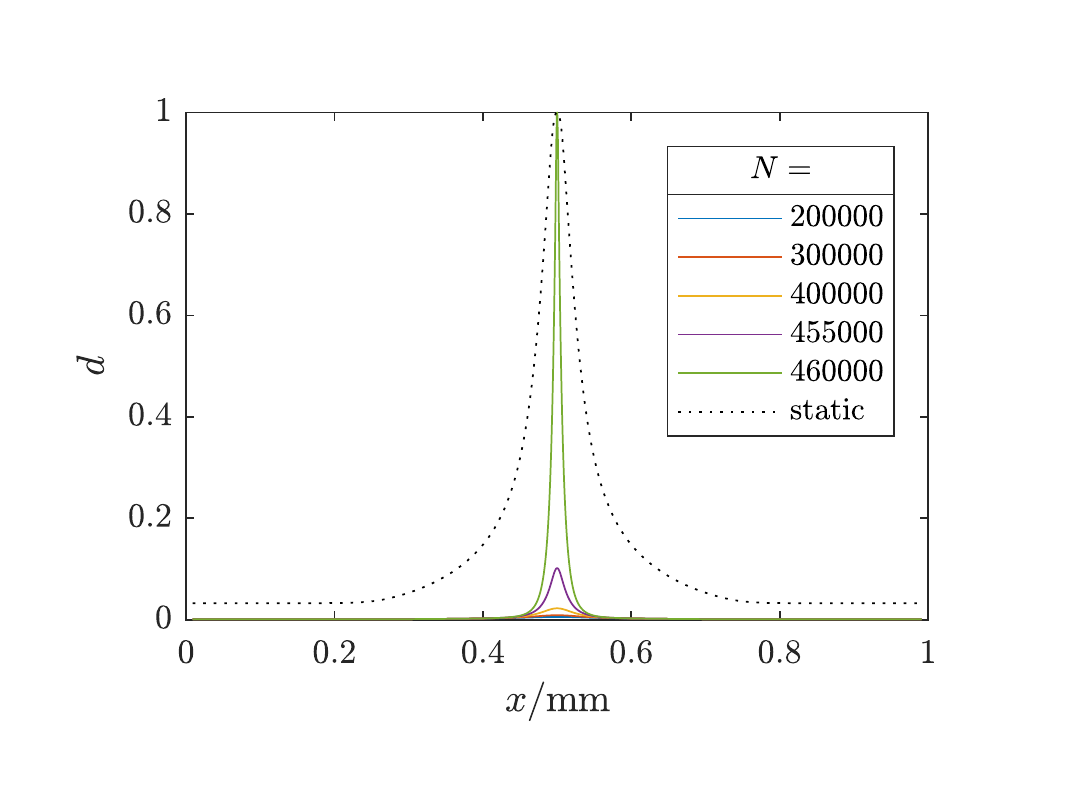} & b)	\\
		\includegraphics[width=0.48\linewidth,trim={0.5cm 0.5cm 1cm 0.5cm},clip] {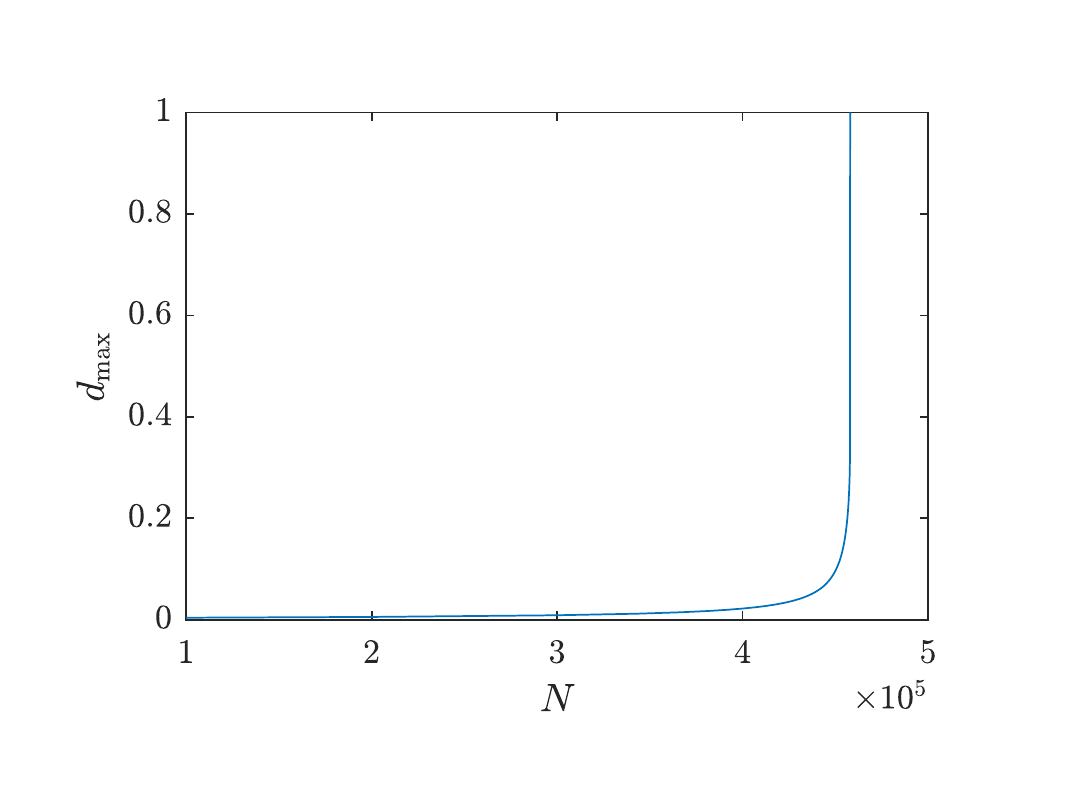}& c) &
		\includegraphics[width=0.48\linewidth,trim={0.5cm 0.5cm 1cm 0.5cm},clip] {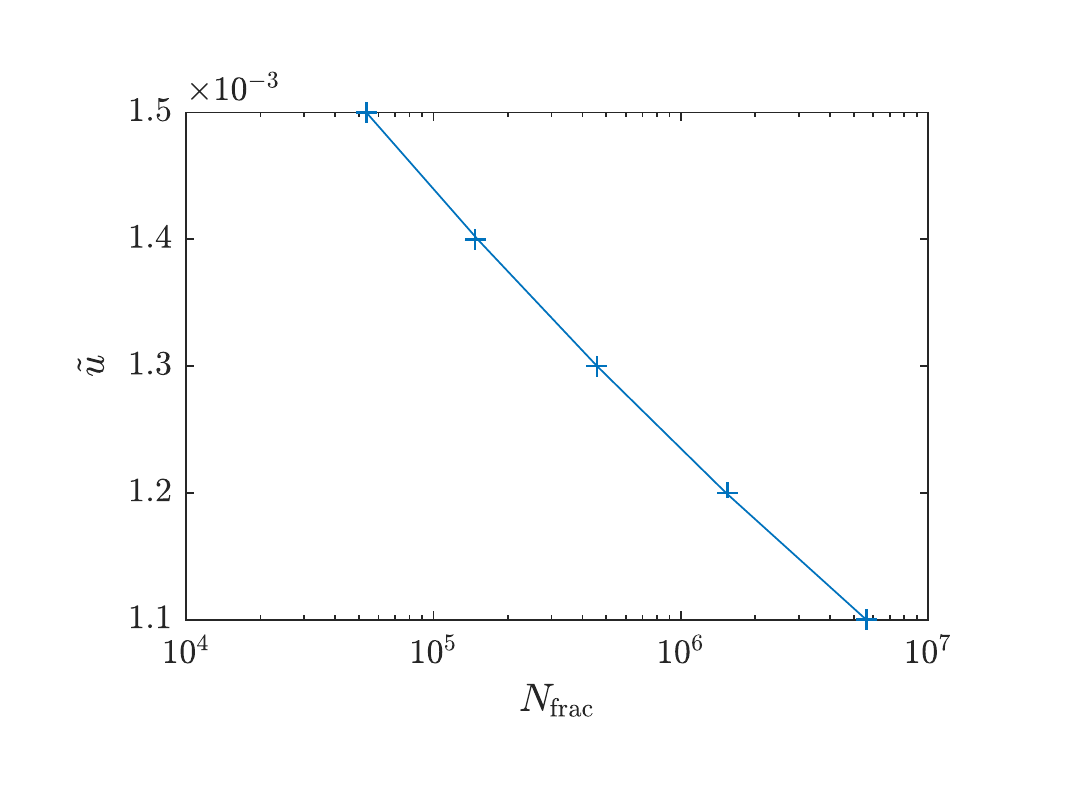} & d)	 \\
		\includegraphics[width=0.48\linewidth,trim={0.5cm 0.5cm 1cm 0.5cm},clip] {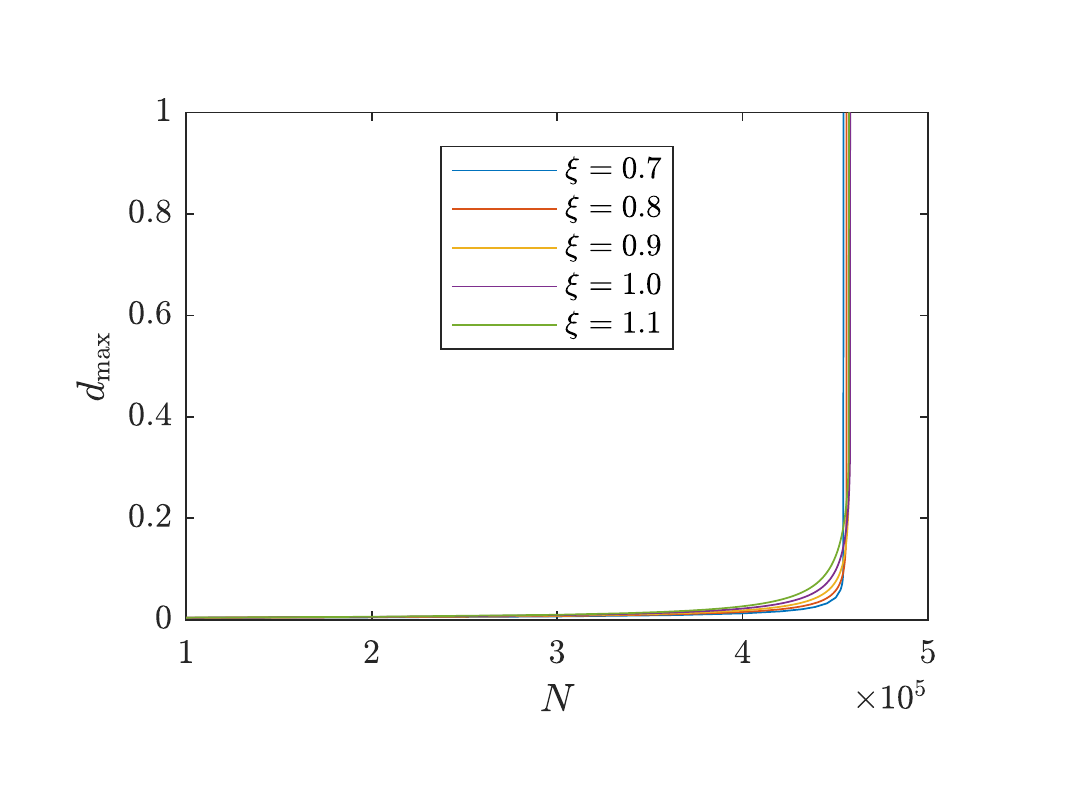} 	& e) &
		\includegraphics[width=0.48\linewidth,trim={0.5cm 0.5cm 1cm 0.5cm},clip] {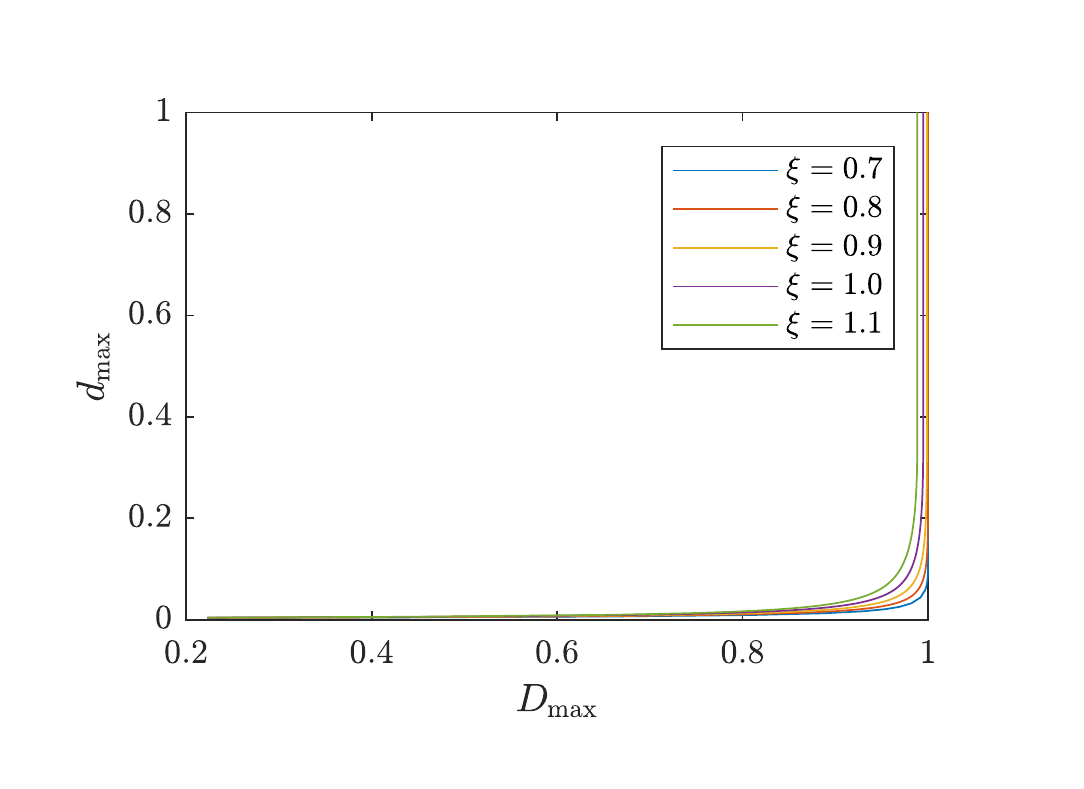} & f) \\ 
	\end{tabular}
	\caption{\textbf{(a)} Evolution of the distribution of the lifetime variable $D$ over the length of the bar $x$ for increasing load cycles $N$. $D$ accumulates in ranges of high stress. 
		\textbf{(b)} Evolution of phase-field $d$. After $N_\mathrm{frac}\approx4.6$ million load cycles, the crack initiates ($d=1$). 
		\textbf{(c)} Evolution of maximum phase-field variable $d_{\max}$ over $N$ shows the successively accelerating damage evolution. 
		\textbf{(d)} Load cycles until fracture $N_\mathrm{frac}$ over displacement load amplitude $\tilde{u}$. The decrease of $N_\mathrm{frac}$ for increasing $\tilde{u}$ can be described by a power function.
		\textbf{(e)} Parameter study for $\xi$ of the fatigue degradation function $\alpha(D)=(1-\alpha_0)(1-D)^\xi+\alpha_0$. 
		Maximum phase-field variable over number of load cycles. The total lifetime is hardly influenced by $\xi$. \textbf{(f)} Relationship of maximum phase-field variable and corresponding lifetime variable, which is primarily controlled by $\xi$.
		\label{fig:NumEx_single}}
\end{figure}

The results of the test are depicted in Fig.~\ref{fig:NumEx_single}. Fig.~\ref{fig:NumEx_single}a) shows the distribution of the lifetime variable $D$ over the length of the bar for increasing load cycles $N$. As expected, $D$ accumulates fastest in the range of the reduced cross section, where stresses and strains are higher and with them the damage parameter $P_\mathrm{SWT}$.  Furthermore, the linear character of $D$ over $N$ becomes obvious, which underlines the meaning of $D$ as expiring lifetime. By comparison, Fig.~\ref{fig:NumEx_single}b) shows the evolution of the phase-field. Obviously, a crack initiates after $N\approx4.6\cdot10^5$ load cycles. The results match observations in W\"{o}hler experiments for metals: Until crack initiation, the component weakens continuously without loosing much stiffness until a crack occurs abruptly. Accordingly, $D$ increases continuously, only degrading the crack resistance. Meanwhile $d$, which degrades the stiffness, develops much slower, until in the end it shows an abrupt increase. This is underlined by Fig.~\ref{fig:NumEx_single}c). For comparison, a phase-field profile of a static simulation with the same material parameters is displayed as well. This shows that the narrow corridor of reduced $\mathcal{G}_\mathrm{c}$ in the fatigue simulation leads to a narrower phase-field profile compared to a static simulation with a homogeneous  $\mathcal{G}_\mathrm{c}$ contribution.

\begin{figure} [t]
	\centering
	\setlength{\tabcolsep}{0pt}
	\begin{tabular}{clcl}
		\includegraphics[width=0.48\linewidth,trim={0.5cm 0.5cm 1cm 0.5cm},clip] {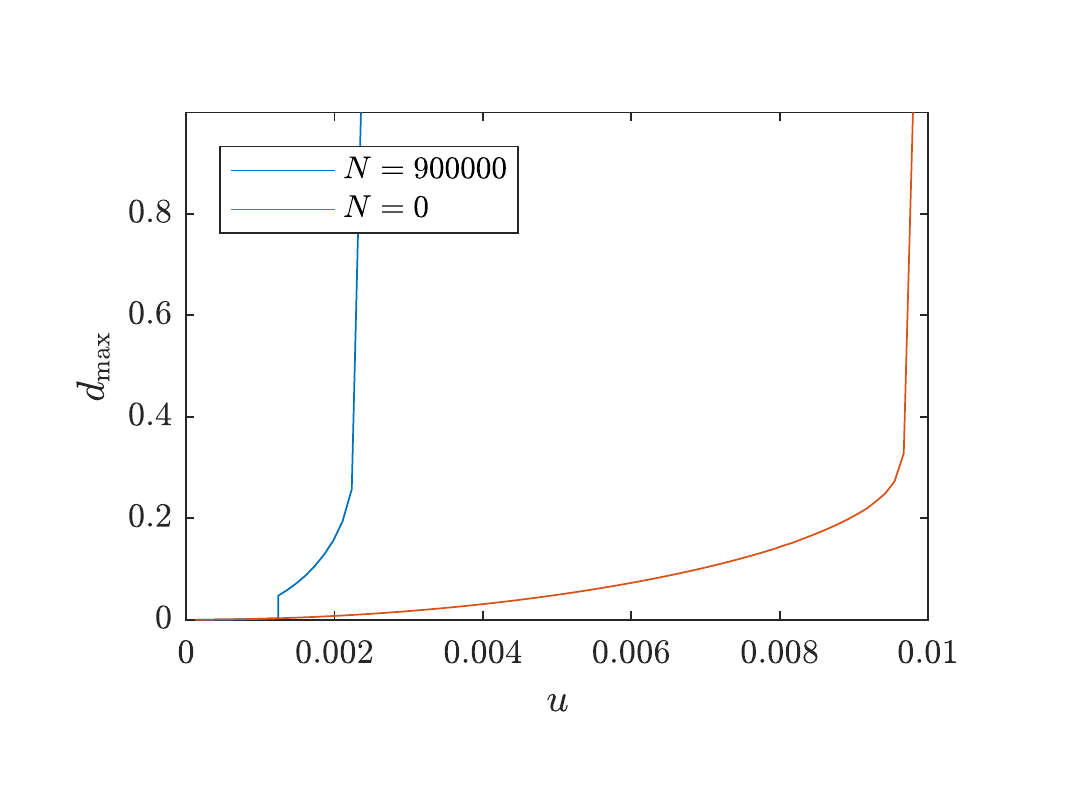} & a) &
		\includegraphics[width=0.48\linewidth,trim={0.5cm 0.5cm 1cm 0.5cm},clip] {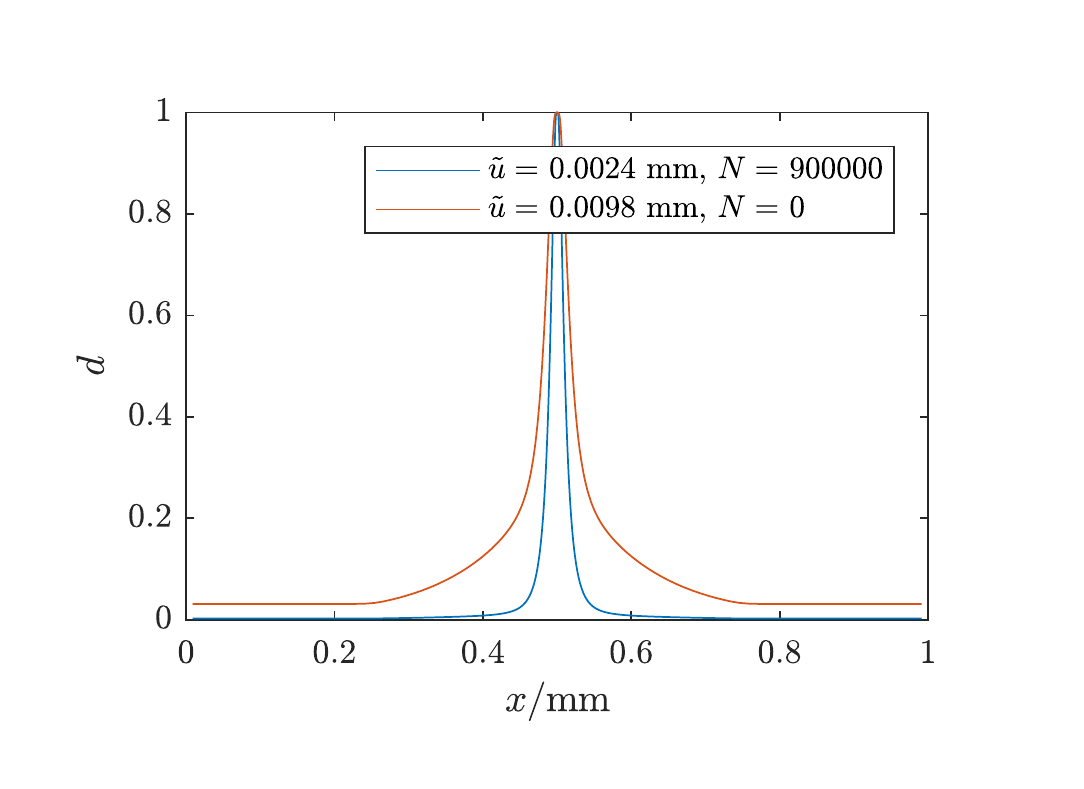} & b)	
	\end{tabular}
	\caption{Two-stage simulation. After $9\cdot10^5$ load cycles (blue line), the load is increased statically, leading to further crack growth under under the influence of the pre-damaged state. A  static simulation with undamaged material (orange line) is displayed for comparison. 
	\textbf{(a)} Maximum phase-field variable $d$ over the length of the bar. \textbf{(b)} Profile of $d$ in the moment of fracture. 	
		\label{fig:NumEx_static}}
\end{figure}

Fig.~\ref{fig:NumEx_single}d) compares the numbers of load cycles until failure $N_\mathrm{frac}$ in the same setting for varying displacement amplitudes $\tilde{u}_\mathrm{a}$. As expected, smaller load amplitudes lead to longer fatigue lives. 
Fig.~\ref{fig:NumEx_single}e) shows the influence of the parameter $\xi$ of the fatigue degradation function $\alpha(D)=(1-\alpha_0)(1-D)^\xi+\alpha_0$. $\xi$ has only a small impact on the total number of load cycles until fracture. Instead, it controls the suddenness of crack formation, which is associated with the brittleness of the material. This link also becomes apparent in Fig.~\ref{fig:NumEx_single}f), where the relation between the phase-field variable and the lifetime variable, both represented by their current maximum $d_{\max}$ and $D_{\max}$, is shown. The higher $\xi$ is, the earlier in terms of the total life and the more gentle the phase-field, i. e. the crack, develops. Generally speaking, the relation between $D$ and $d$ cannot be expressed by a function $d(D)$. The development of the crack field e.\,g. also depends on the current stress-strain state and the phase-field gradient $\nabla d$. This becomes clear in a fatigue test with a subsequent static load as shown in Fig.~\ref{fig:NumEx_static}. After $9\cdot10^5$ load cycles at a displacement amplitude of $\tilde{u}_\mathrm{a}=0.0012$\,mm, a static load is applied. Due to the prior damage by the fatigue load the crack initiates at a displacement which is only one fourth compared to undamaged material. Interestingly, the phase-field shows a more narrow profile due to the inhomogeneous distribution of the fracture toughness.

\subsection{Two-dimensional examples}
\label{sec:NumEx_2D}

\subsubsection{Single-edge notched test}
\label{sec:SEN}

In a two-dimensional setting, the method is at first tested with the single-edge notched specimen subject to tensile and shear loading. 
The characteristic length is chosen to $\ell=0.01\,L$ with regard to the edge length $L=100$ mm.
A hierarchically refined quadrilateral mesh with 8 element size levels and quadratic ansatz functions is applied \cite{hennig_bezier_2016}. Within the area of the growing crack the minimum element size is  $h_\mathrm{min}=1/3.2\,\ell$. Young's modulus, Poisson's ratio and fracture toughness are chosen to $E=210$ GPa, $\nu=0.3$ and $\mathcal{G}_\mathrm{c}=2.7$ N/mm. The parameters of the fatigue degradation function are set to $\alpha_0=0.05$ and $\xi=1$. No split is applied to the energy density. The load cycles simulated by one increment $\Delta N$ are adjusted according to the required number of staggered loops. The specimen is subject to plane strain condition and loaded with a constant displacement amplitude of $\tilde{u}=0.0018\,L$ in the tension test and $\tilde{u}=0.003\,L$ in the shear test.

\begin{figure} [p]
	\setlength{\tabcolsep}{-2pt}
	\begin{tabular}{m{3.8cm}m{0.8cm}m{0.5cm}m{1.2cm}m{3.8cm}m{3.8cm}m{3.8cm}m{0cm}}
		\centering static & &&& \centering $N=6000$  & \centering $N=14600$ &\centering $N=15360$&
		\\ 
		\includegraphics[width=\linewidth]{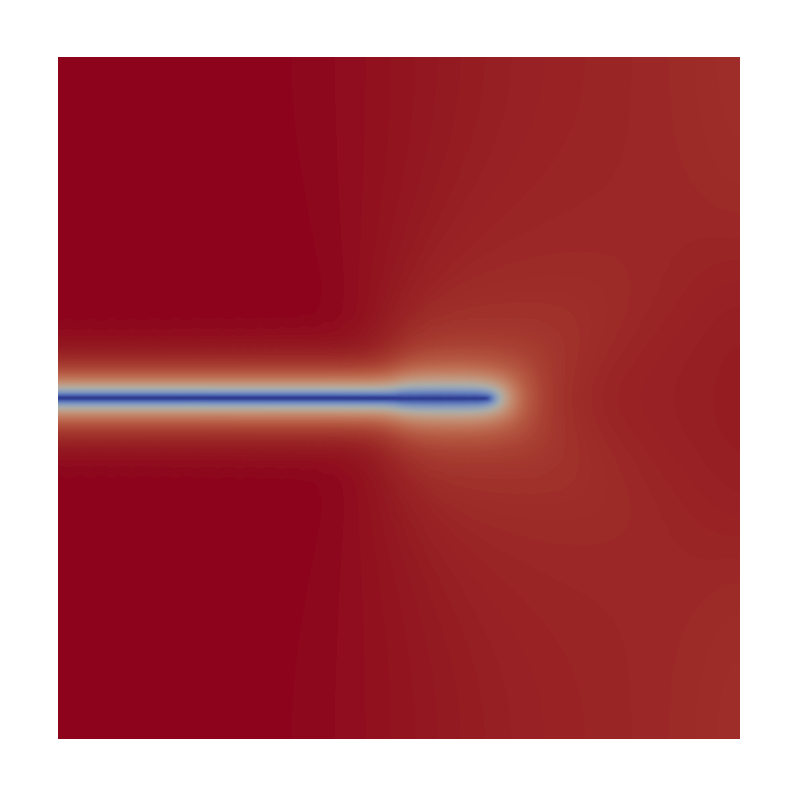} &&$d$ &
		\includegraphics[width=\linewidth]{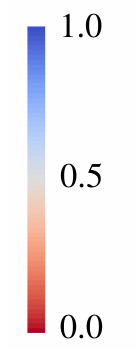}&
		\includegraphics[width=\linewidth]{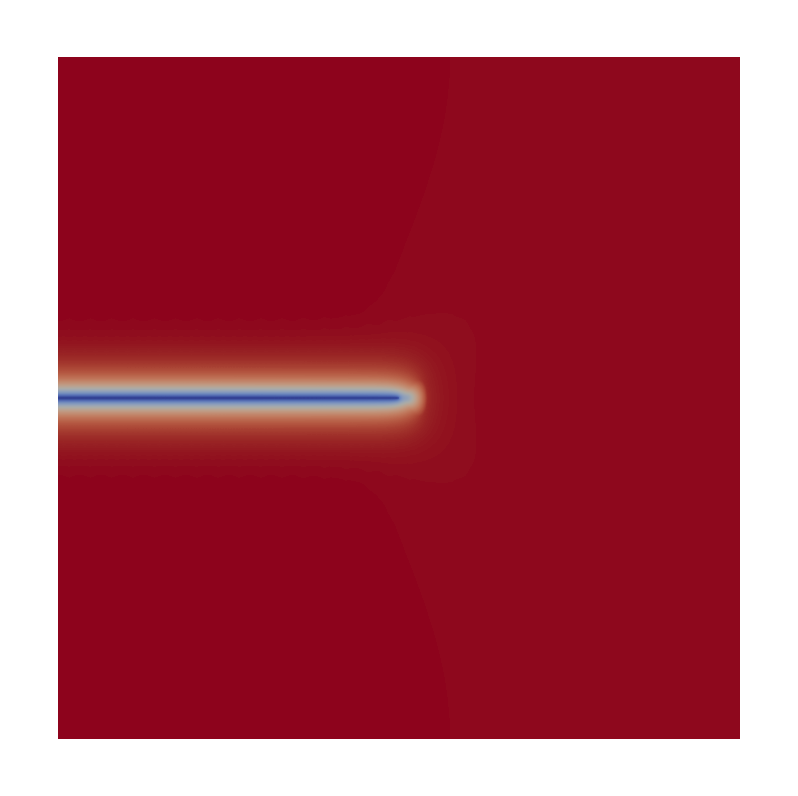} &   
		\includegraphics[width=\linewidth]{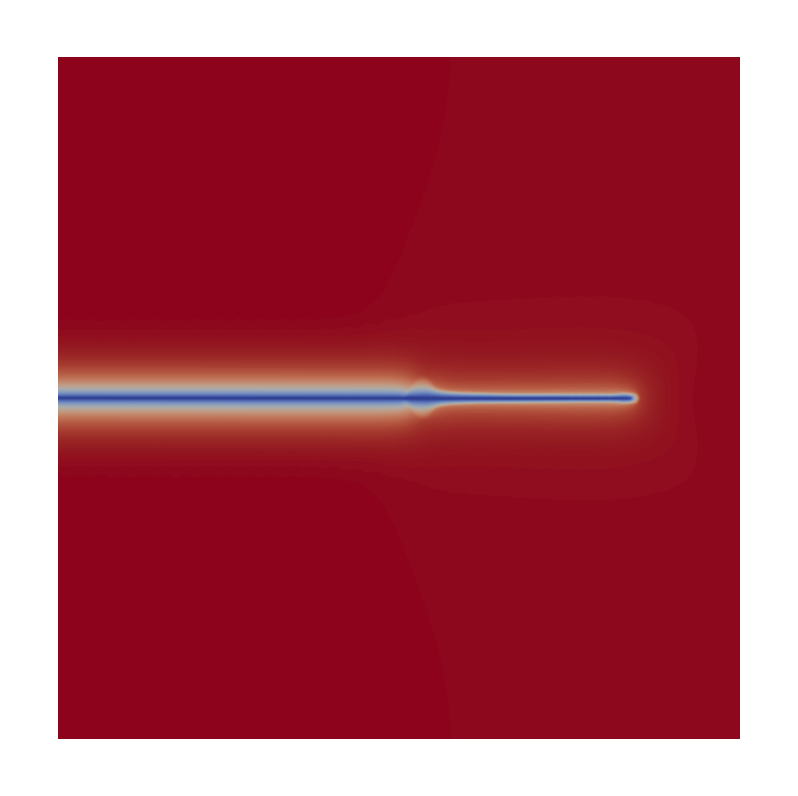} &
		\includegraphics[width=\linewidth]{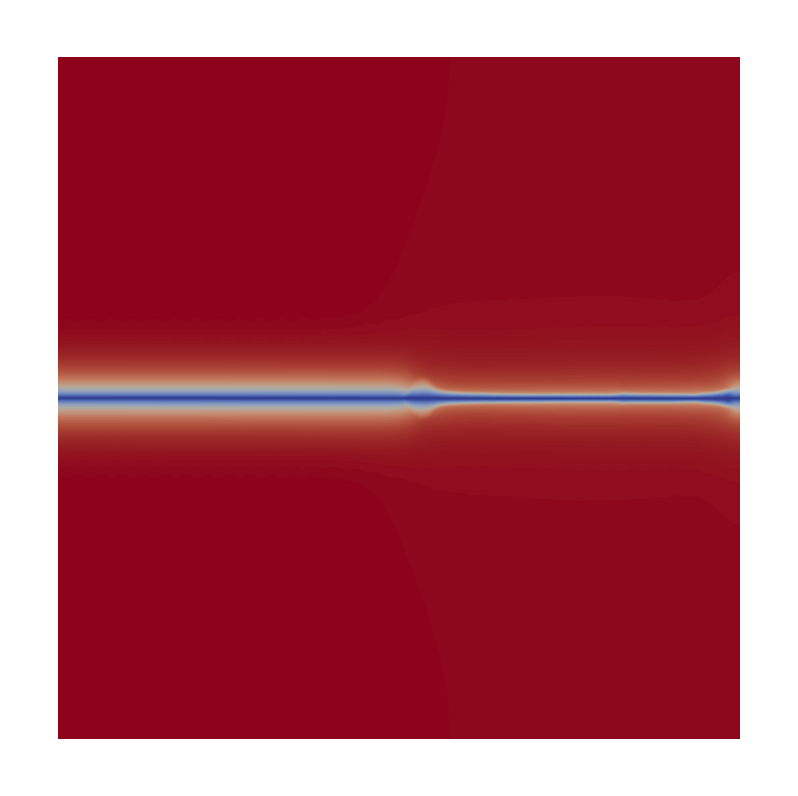} &
		\\
		\includegraphics[width=\linewidth]{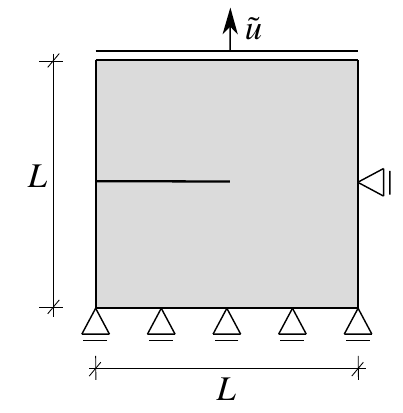}& &$D$ &
		\includegraphics[width=\linewidth]{Pics/2D/scale_vertical/pdfgen.pdf}&
		\includegraphics[width=\linewidth]{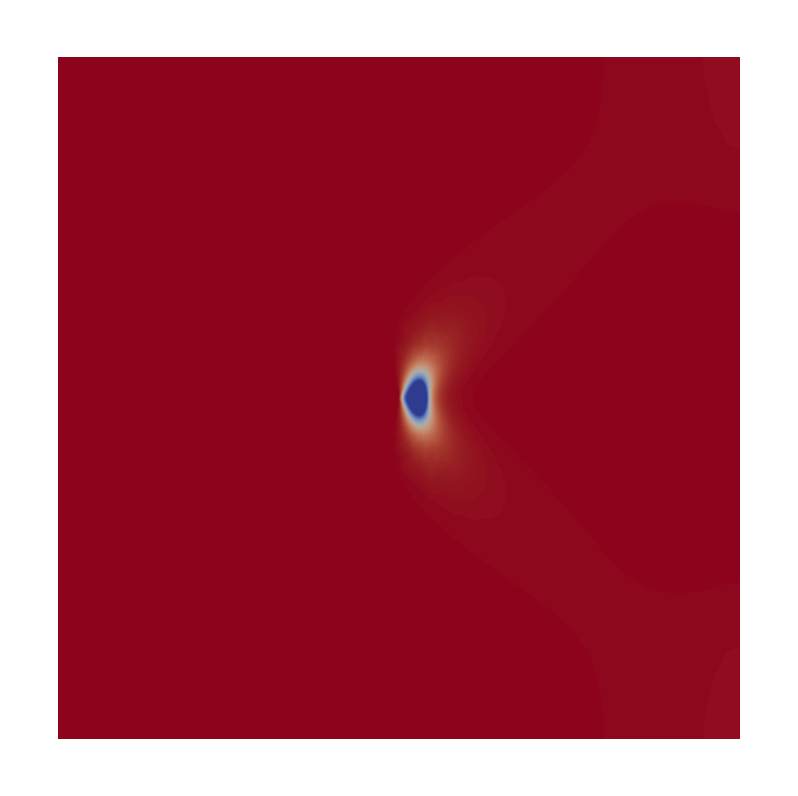} &   
		\includegraphics[width=\linewidth]{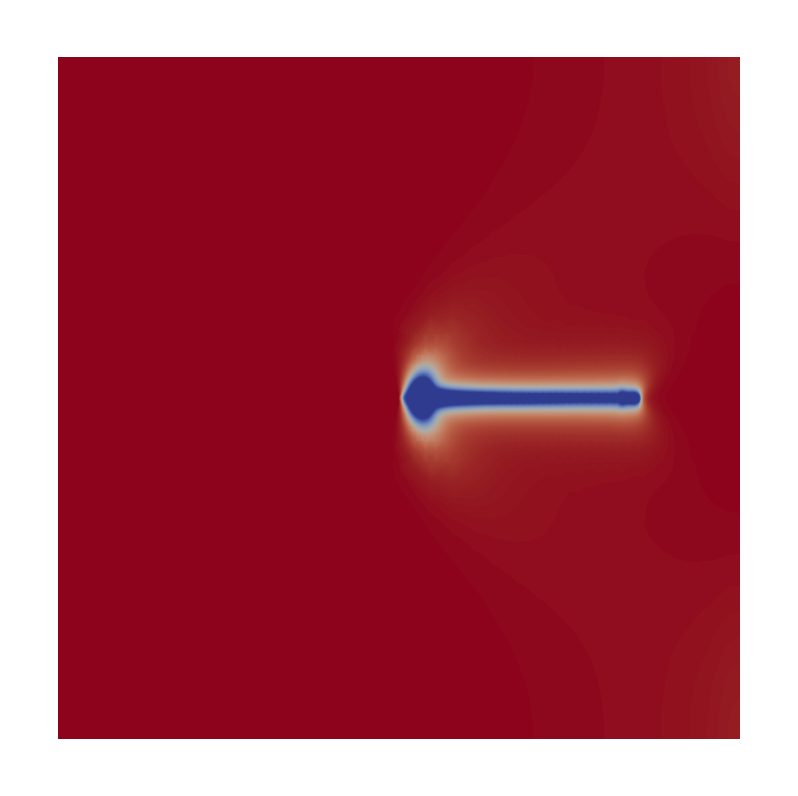} &
		\includegraphics[width=\linewidth]{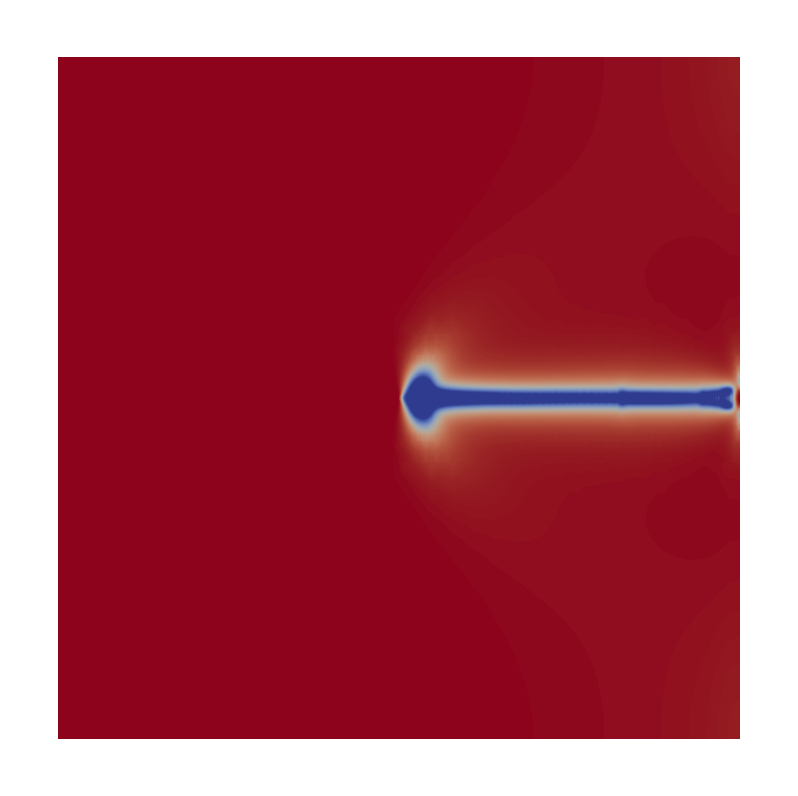} & 
	\end{tabular} 			 
	\caption{Single-edge notched \textbf{\textit{tension}} test ($L=100$ mm) with cyclic loading. Distribution of phase-field $d$ and lifetime variable $D$ for different numbers of load cycles $N$. Static case with same characteristic length $\ell=0.01\,L$ for comparison.
		\label{fig:NumEx_SENten}}
\end{figure}
\begin{figure} [p]
	\setlength{\tabcolsep}{-2pt}
	\begin{tabular}{m{3.8cm}m{0.8cm}m{0.5cm}m{1.2cm}m{3.8cm}m{3.8cm}m{3.8cm}m{0cm}}
		\centering static && && \centering $N=3600$ & \centering $N=13500$ &\centering $N=15330$&
		\\
		\includegraphics[width=\linewidth]{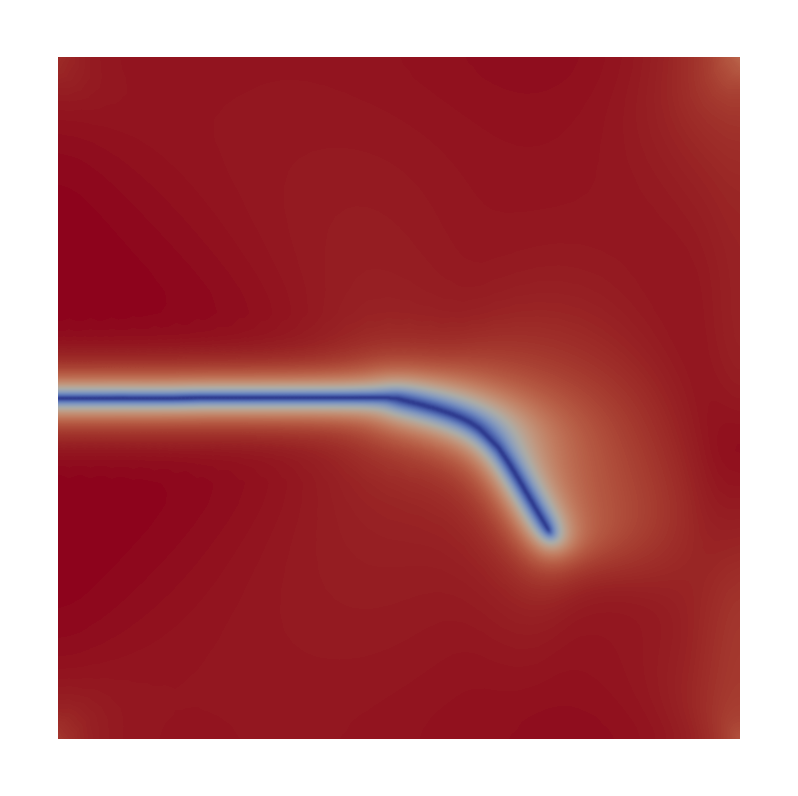} &&$d$ &
		\includegraphics[width=\linewidth]{Pics/2D/scale_vertical/pdfgen.pdf}&
		\includegraphics[width=\linewidth]{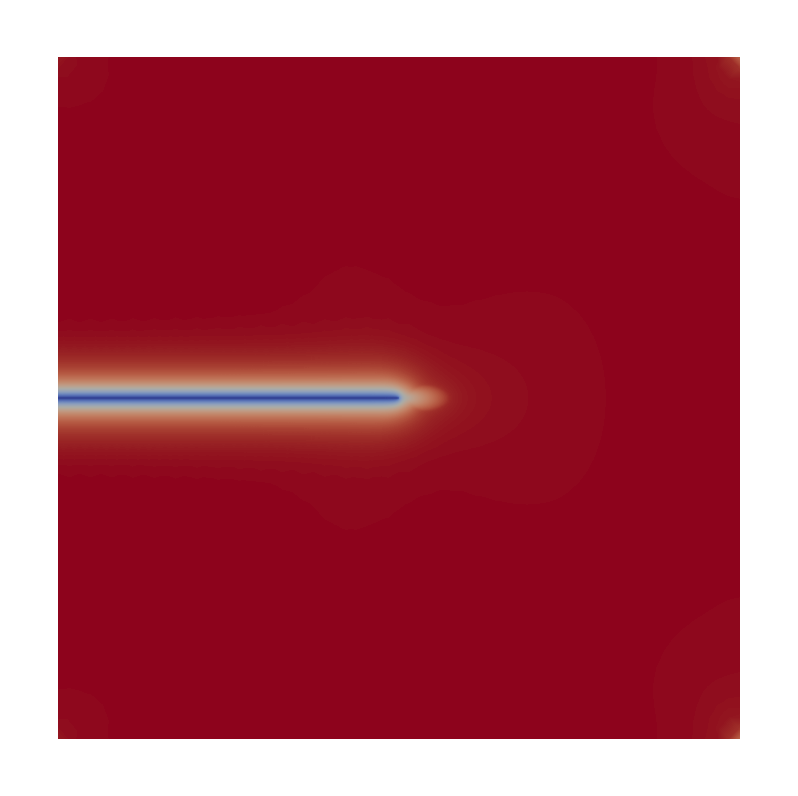} & 
		\includegraphics[width=\linewidth]{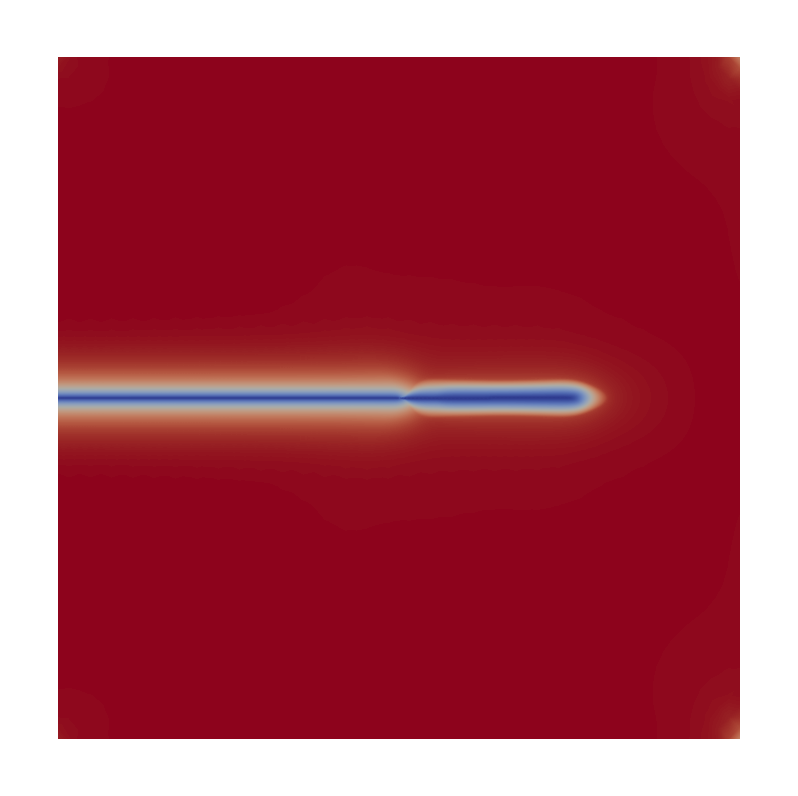} & 
		\includegraphics[width=\linewidth]{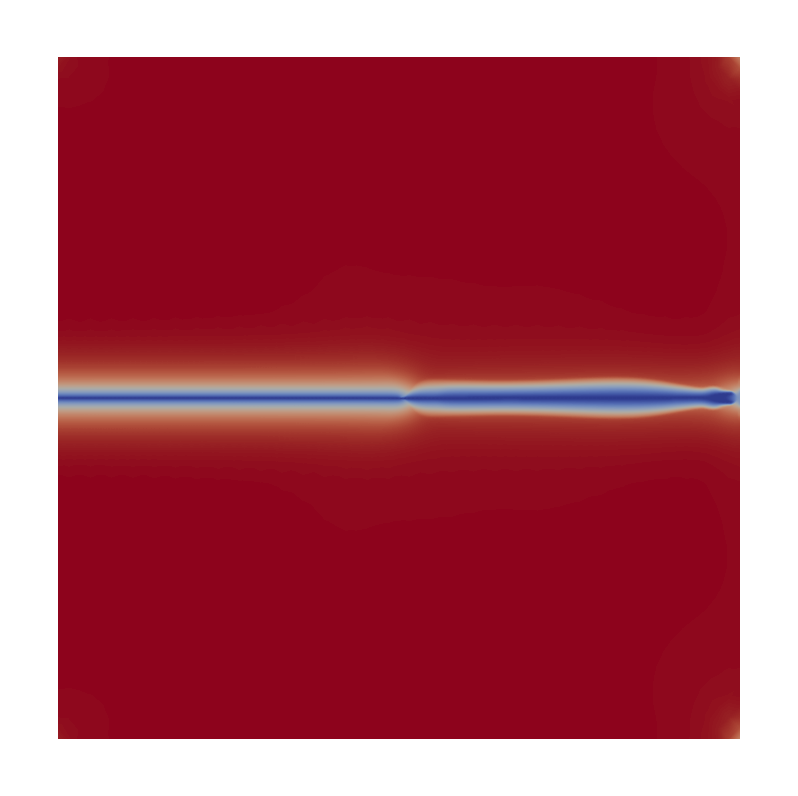} &
		\\
		\includegraphics[width=\linewidth]{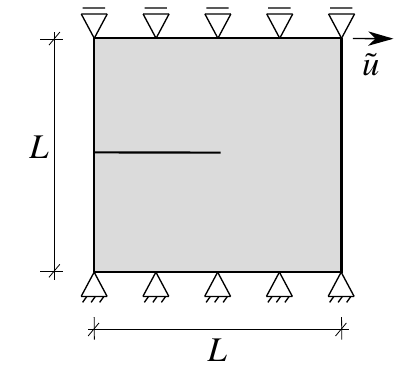} &&
		$D$ &\includegraphics[width=\linewidth]{Pics/2D/scale_vertical/pdfgen.pdf}&
		\includegraphics[width=\linewidth]{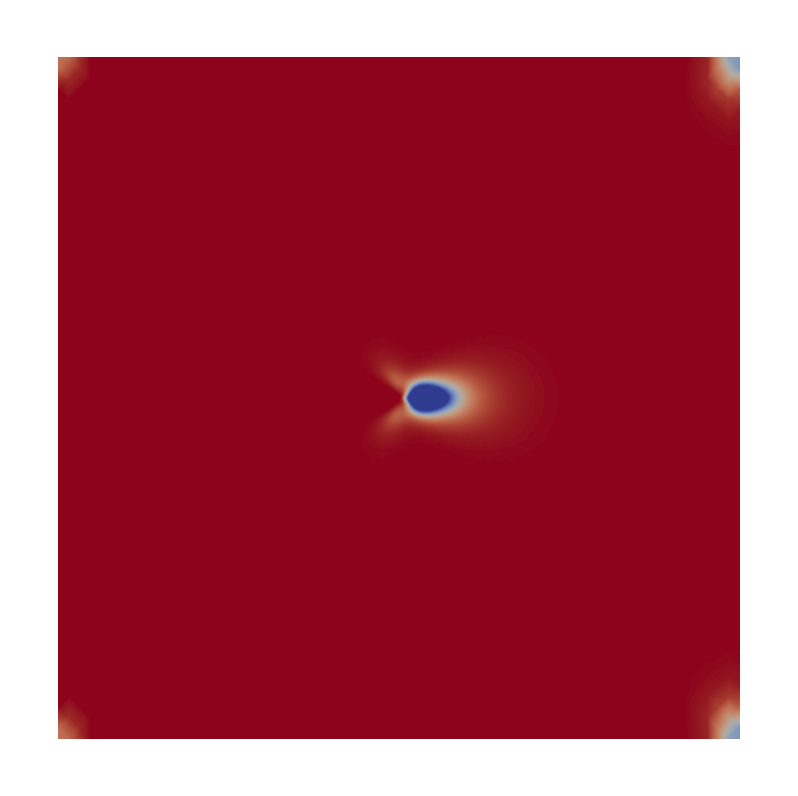} & 
		\includegraphics[width=\linewidth]{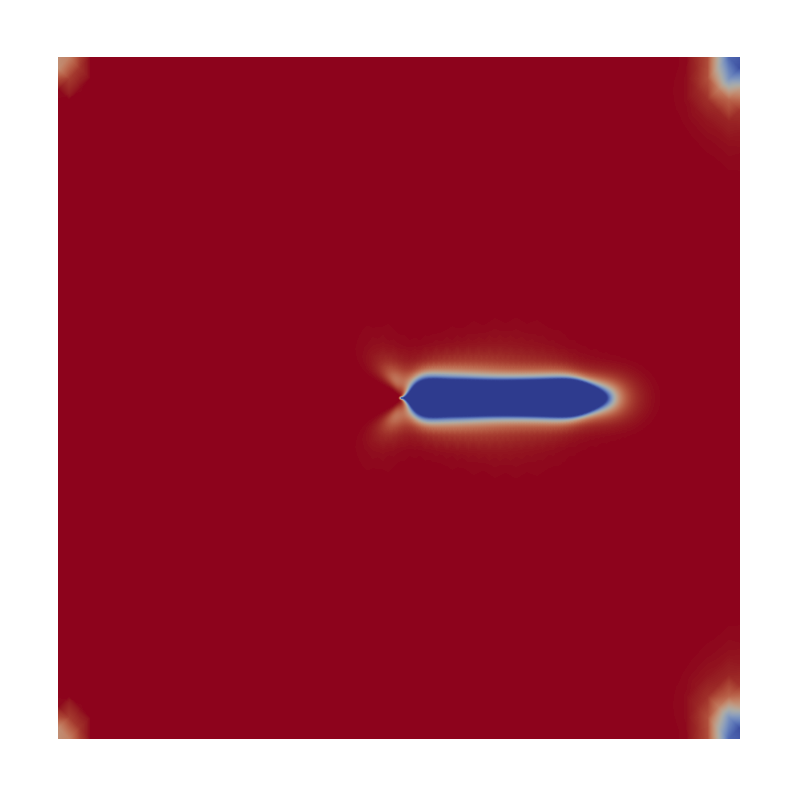} & 
		\includegraphics[width=\linewidth]{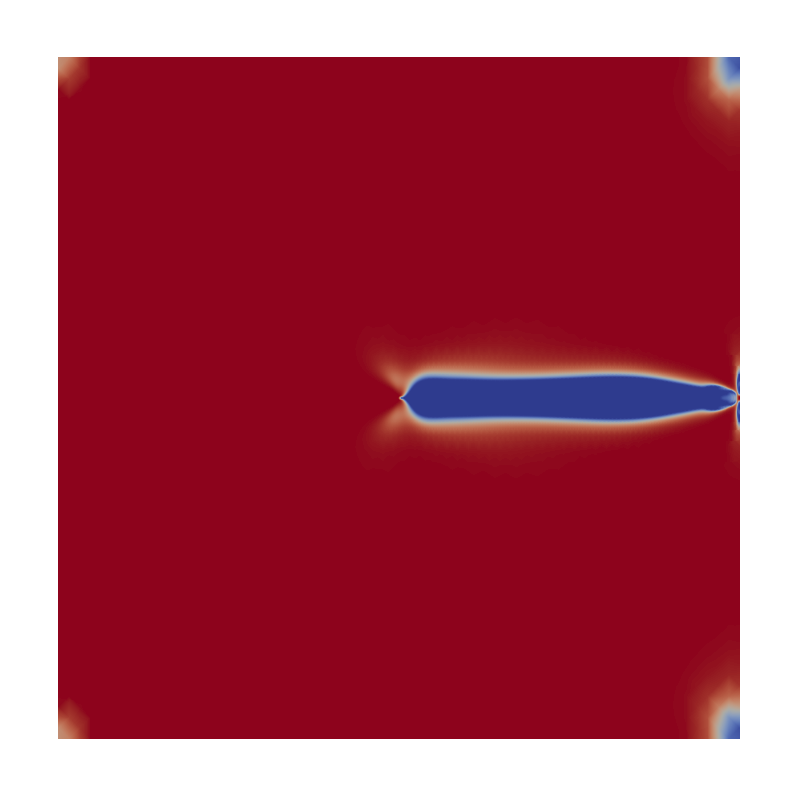} & 
	\end{tabular} 	 
	\caption{Single-edge notched \textbf{\textit{shear}} test ($L=100$ mm) with cyclic loading. Distribution of phase-field $d$ and lifetime variable $D$ for different numbers of load cycles $N$.  Static case with same characteristic length $\ell=0.01\,L$ and Amor split \cite{amor_regularized_2009} for comparison.
		\label{fig:NumEx_SENshear}}
\end{figure}

The evolution of the phase-field and the lifetime variable are displayed in Fig.~\ref{fig:NumEx_SENten} (tension) and Fig.~\ref{fig:NumEx_SENshear} (shear). Starting from the notch, a zone of $D=1$ forms progressively. This area can be associated with the plastic zone, since $D$ is computed from the damage parameter $P_\mathrm{SWT}$, which is again associated with the area inside a stress-strain hysteresis -- a measure for dissipation. The crack forms in the corridor of material with reduced fracture toughness. It shows a more narrow phase-field profile compared to the static case with the same characteristic length $\ell$ also displayed in Fig.~\ref{fig:NumEx_SENten}. Since $\alpha(D)$ is a factor in the gradient term of the phase-field  equation (\ref{eq:ev_PF}), which controls the profile of the phase-field crack, a deviation from the profile of a static simulation with a homogeneous distribution of $\mathcal{G}_\mathrm{c}$ is expected.

Interestingly, in the shear test the crack runs straight through the specimen. For a static, brittle phase-field formulation with a tension-compression split -- also displayed in Fig.~\ref{fig:NumEx_SENshear} -- the crack is deflected towards the upper or lower edge depending on the load direction. For a cyclic, brittle formulation and alternating load with $\tilde{u}_{\min}=-\tilde{u}_{\max}$ a branching crack is expected even with a tension-compression split: Due to the changing load direction crack propagation occurs at the top and bottom crack front alternately \cite{carrara_novel_2018}. 
The model for fatigue fracture presented here does not involve a split. It can therefore consider the reversion of the stress state during the load cycle although only the top load is applied to the specimen. Anyway, the crack forms straight due to the ductile formulation of this model: The reduction of $\mathcal{G}_\mathrm{c}$ is derived from the equivalent stress $\sigma_\mathrm{eq}$ which is the highest horizontally in front of the crack tip. This leads the crack to propagate horizontally in this case. The crack path also corresponds to results for a ductile phase-field model by Miehe et al. \cite{miehe_phase_2016}, where a single-edge notched specimen is subjected to static shear loading.

\subsubsection{Compact tension test}

\begin{figure} [t]
	\centering
	\setlength{\tabcolsep}{0pt}
	\begin{tabular}{clclcl}
	\includegraphics[width=0.29\linewidth]{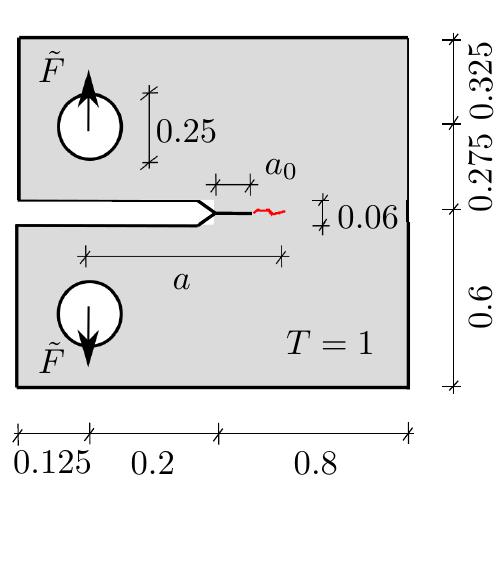} 	& a) & \includegraphics[width=0.35\linewidth]{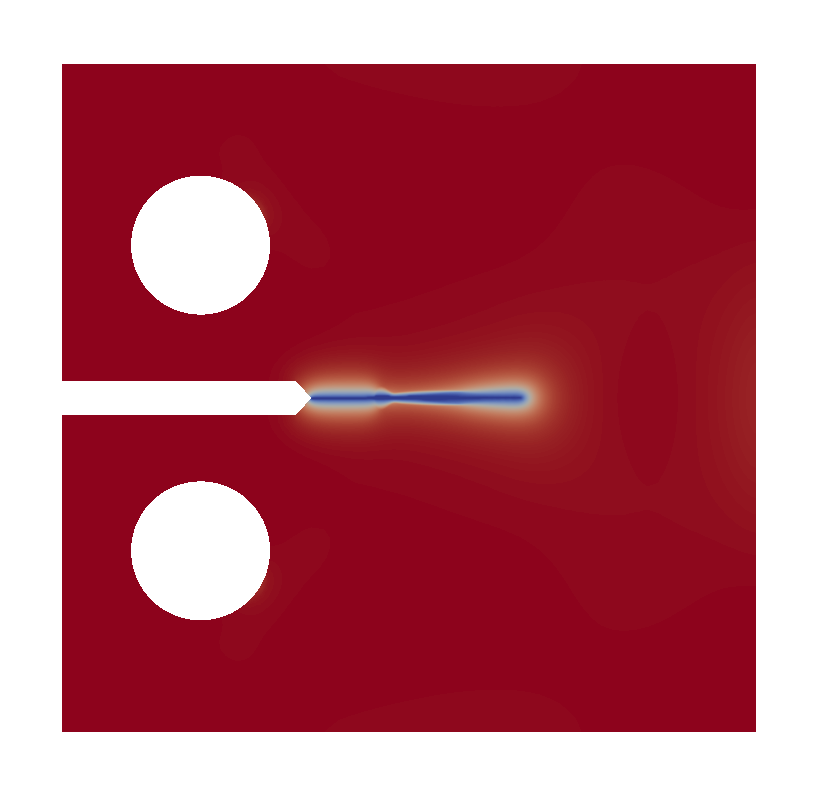} 	& b) & \includegraphics[width=0.28\linewidth]{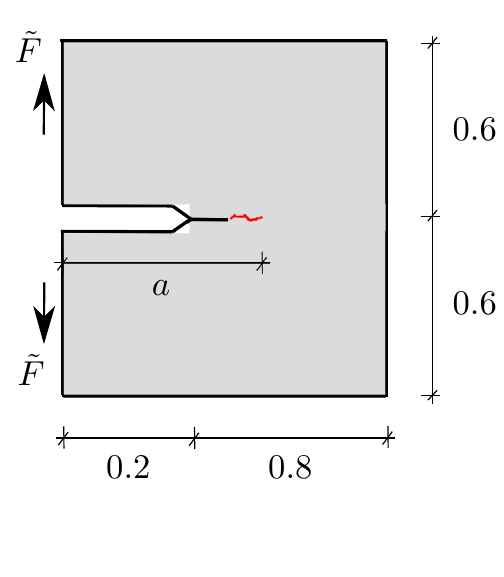}  & c)
	\end{tabular}	 
	\caption{\textbf{(a)} Setup of Compact tension (CT) test according to guidelines ASTM E647-05 \cite{noauthor_standard_2005} and ASTM E1820-01  \cite{noauthor_standard_2001} All dimensions are given in $1/W$. \textbf{(b)} Cyclic simulation of CT test with $a_0=0.1\,W$ and $\tilde{F}=0.07$ kN: Phase-field distribution $d$ after $N=3790$ load cycles. \textbf{(c)} Simplified model. 
		\label{fig:CT_setup}}
\end{figure}

In order to study the Paris behaviour of the method, a compact tension test is considered.
Fig.~\ref{fig:CT_setup}a) displays the setting of the test which follows the guidelines ASTM E647-05 \cite{noauthor_standard_2005} and ASTM E1820-01  \cite{noauthor_standard_2001}. 
The specimen is loaded with a constant force amplitude of $\tilde{F}=0.11$ kN and no mean load, which again entails a constant force boundary condition for this approach. The initial crack length is set to $a_0=0.1\,W$. The same mesh refinement and model parameters as in the single-edge notched specimen (Section~\ref{sec:SEN}) are applied, if not stated differently. The load cycles simulated by one increment are again controlled adaptively by the required staggered loops. Any point with $d>0.95$ is considered a crack.

\begin{figure} [h]
	\centering
	\setlength{\tabcolsep}{0pt}
	\begin{tabular}{clcl}
		\includegraphics[width=0.48\linewidth]{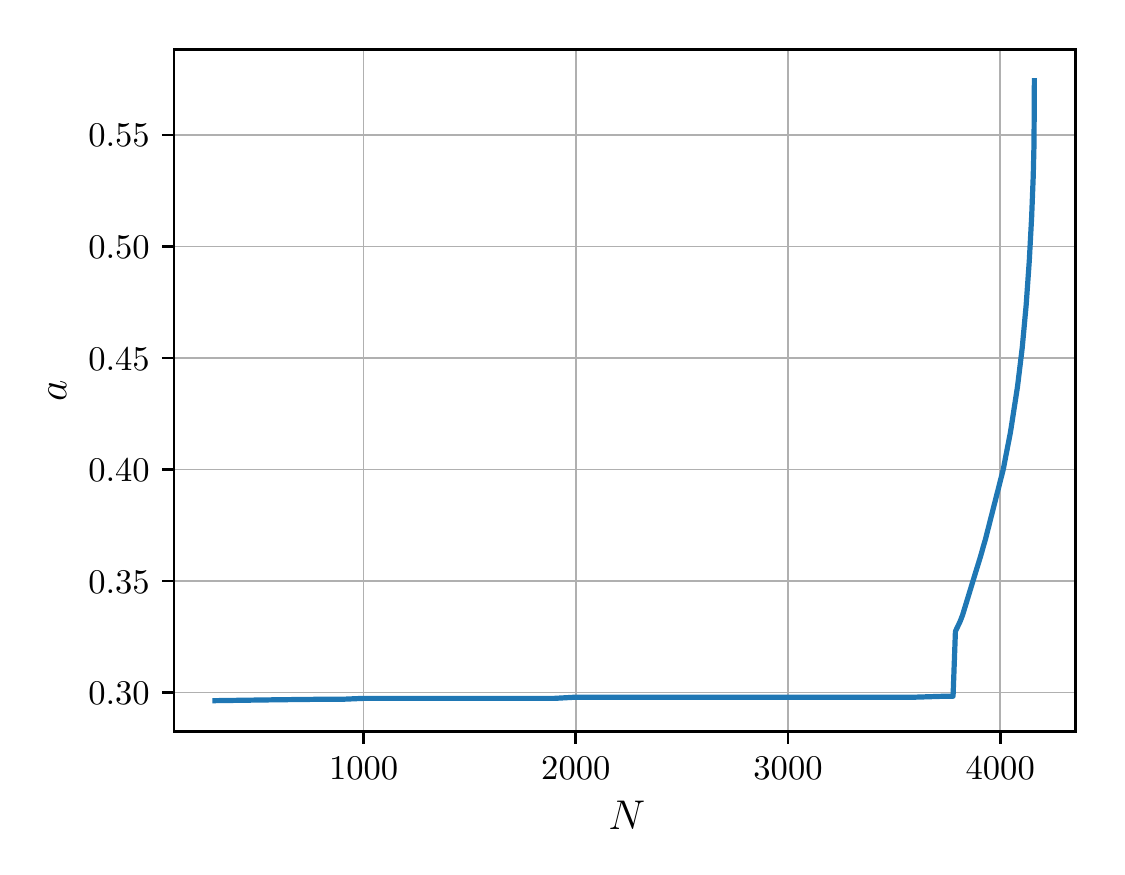} & a) &
		\includegraphics[width=0.48\linewidth]{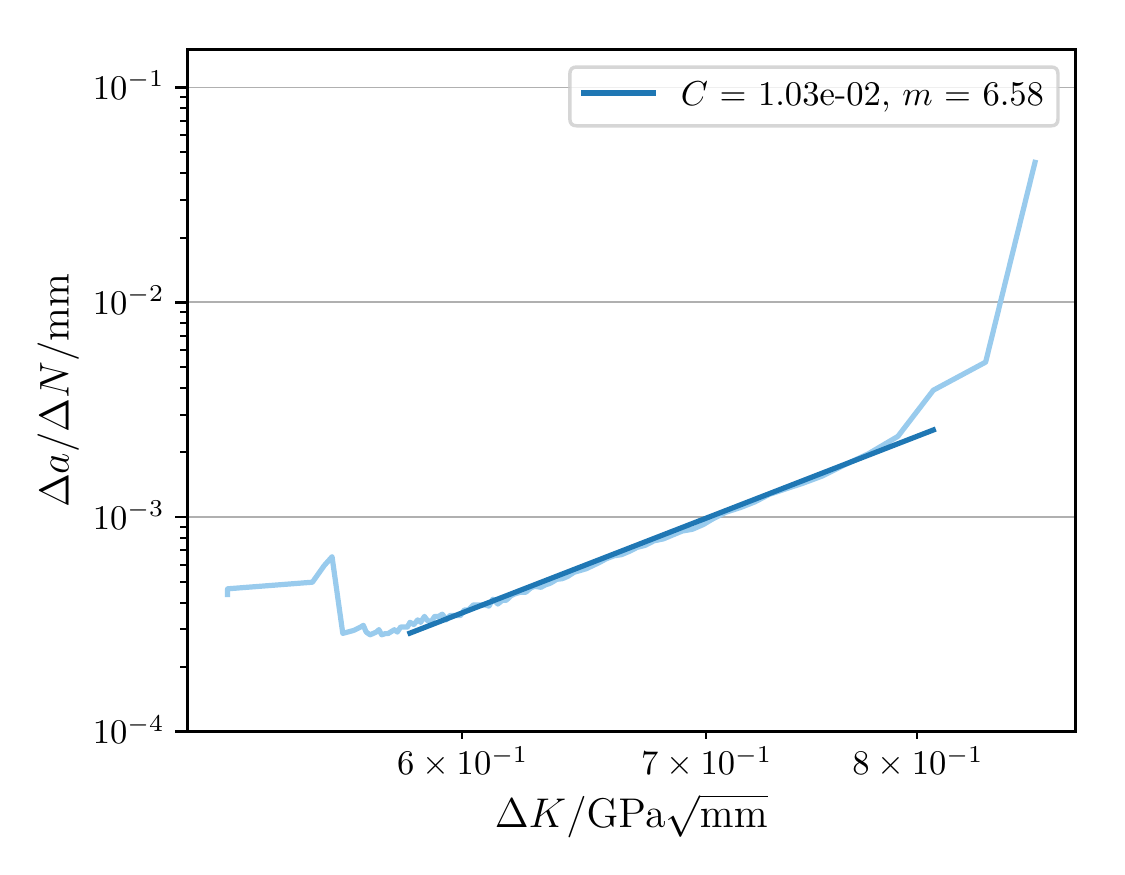} & b)\\
		\includegraphics[width=0.48\linewidth]{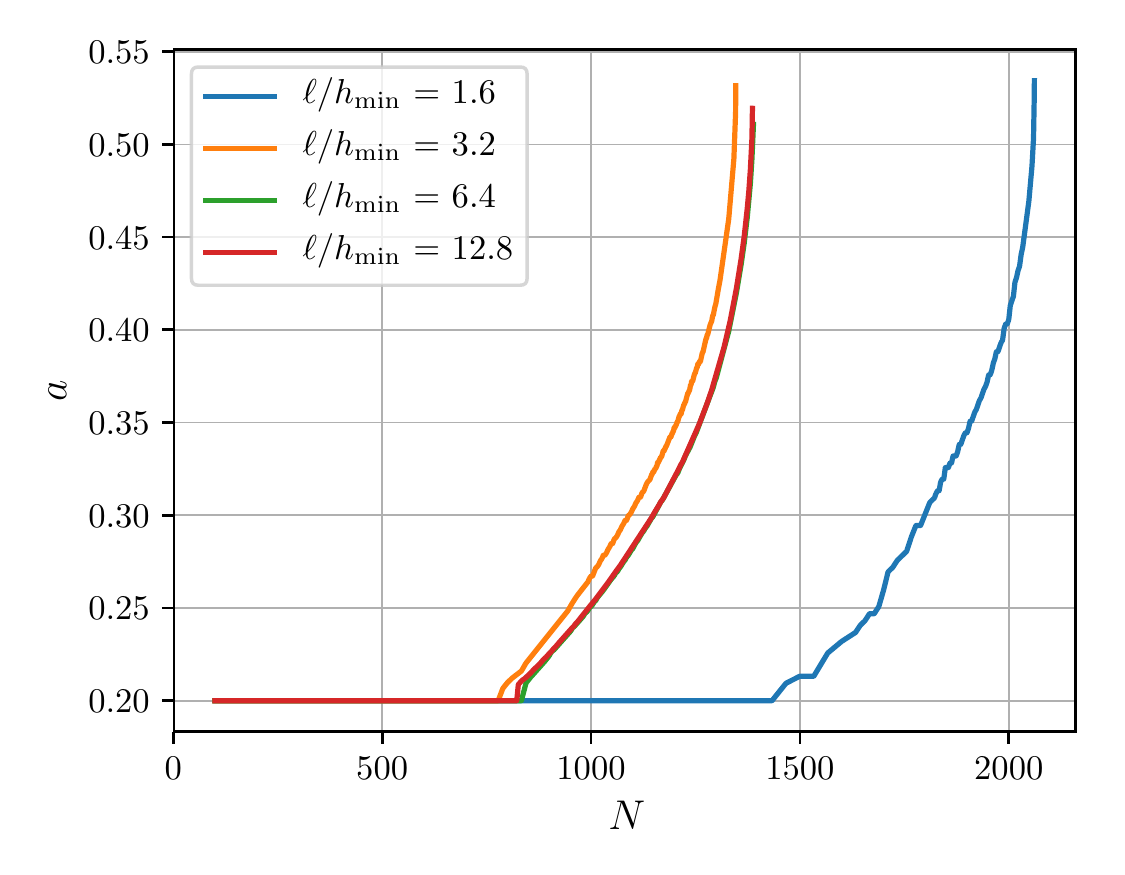} & c) &
		\includegraphics[width=0.48\linewidth]{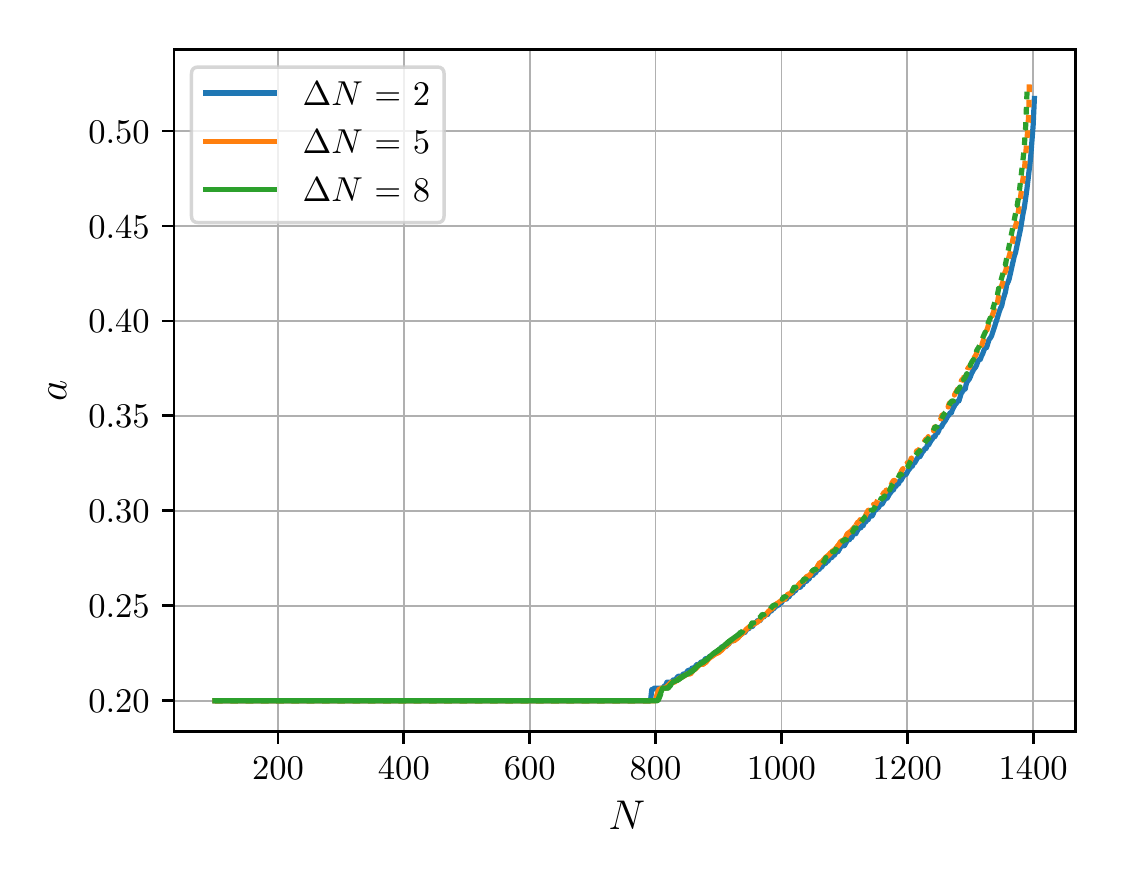} & d)\\
		\includegraphics[width=0.48\linewidth]{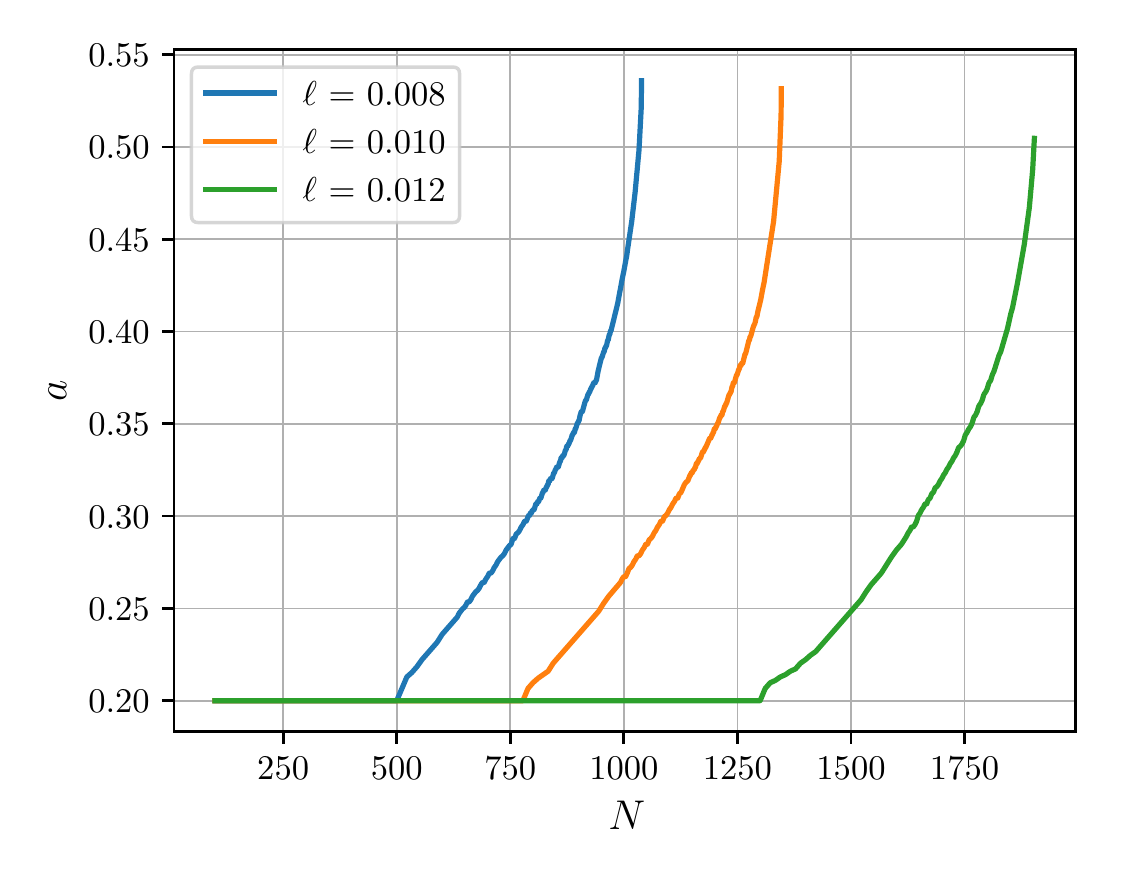} & e) &
		\includegraphics[width=0.48\linewidth]{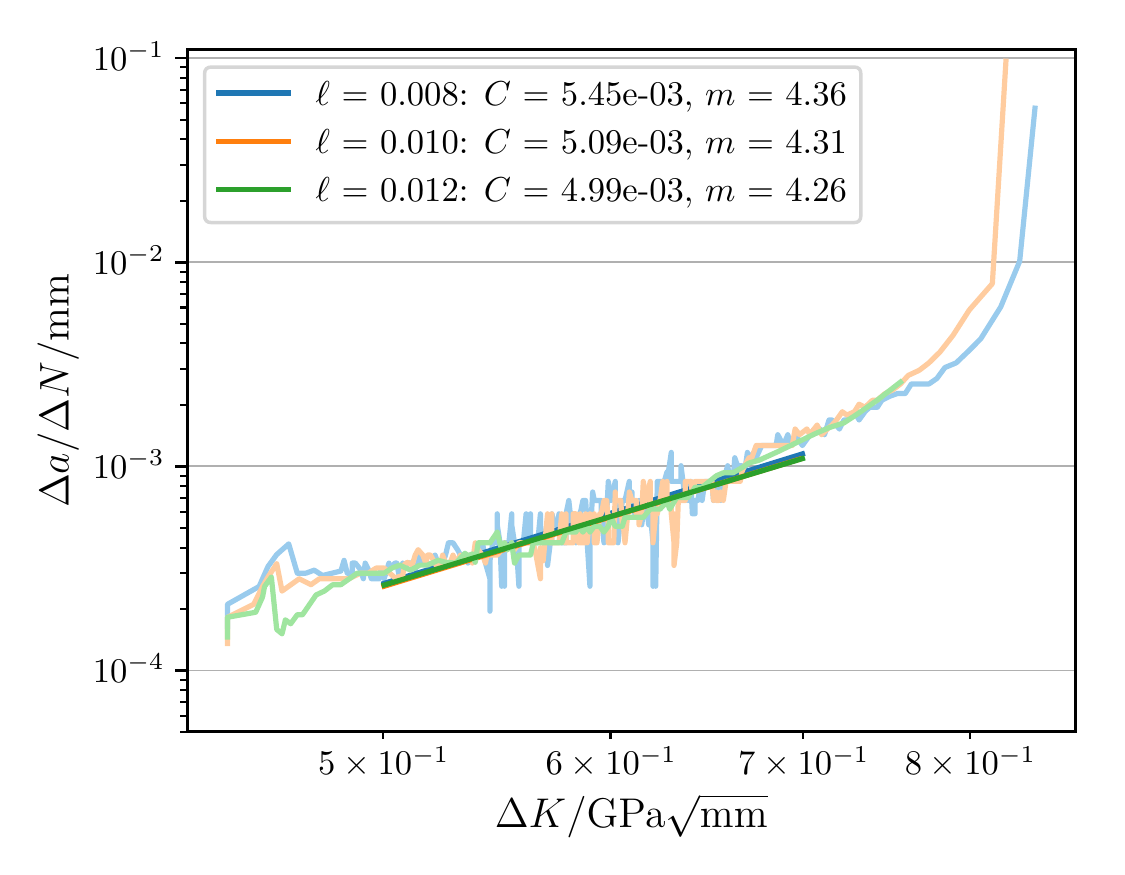} & f)
	\end{tabular}			 
	\caption{Cyclic compact tension test. Parameters: element size $h_\mathrm{min}$, load cycles per increment $\Delta N$, characteristic length $\ell$ in mm. \textbf{(a)}, \textbf{(c)}, \textbf{(d)}, \textbf{(e)} Crack length $a$ over number of load cycles $N$. \textbf{(b)}, \textbf{(f)} Paris plot: Crack propagation rate $\Delta a/\Delta N$ over amplitude of stress intensity factor at crack tip $\Delta K$. Fitted with the Paris parameters $C$ and $m$ within the Paris range. In \textbf{(d)} the dashed line marks the stage of the simulation when $\Delta N$ was reduced the given value.
		\label{fig:2D1}}
\end{figure}

Fig.~\ref{fig:CT_setup}b) shows the crack after $N=3790$ load cycles. The resulting crack length $a$ is displayed in Fig.~\ref{fig:2D1}a). After $\approx3790$ load cycles, a crack initiates and then propagates with an increasing crack propagation rate $\Delta a/\Delta N$. This rate is plotted over the amplitude of the stress intensity factor $\Delta K$ in a Paris plot in Fig.~\ref{fig:2D1}b). Although $\Delta K$ is only valid for small plastic zones, it is here used for the sake of comparability. For this geometry, $\Delta K$ is derived from the crack length $a$ according to the mentioned ASTM guidelines 
\begin{equation}
\Delta K = \frac{\tilde{F}}{T\sqrt{W}}\frac{2+a/W}{(1-a/W)^{3/2}} \left( 0.866 + 4.64 \frac{a}{W} - 13.32 \left(\frac{a}{W}\right)^2 +14.72\left(\frac{a}{W}\right)^3 -5.6 \left(\frac{a}{W}\right)^4 \right).
\end{equation}
After the stable crack propagation stage, to which the Paris parameters $C$ and $m$ of Eq.~(\ref{eq:Paris}) are fitted, the crack proceeds instablely. In the contour plot  Fig.~\ref{fig:CT_setup}b) this residual crack then shows a smoother profile similar to the initial (static) crack.
\begin{figure} [h]
	\centering
	\setlength{\tabcolsep}{0pt}
	\begin{tabular}{clcl}
		\includegraphics[width=0.5\linewidth]{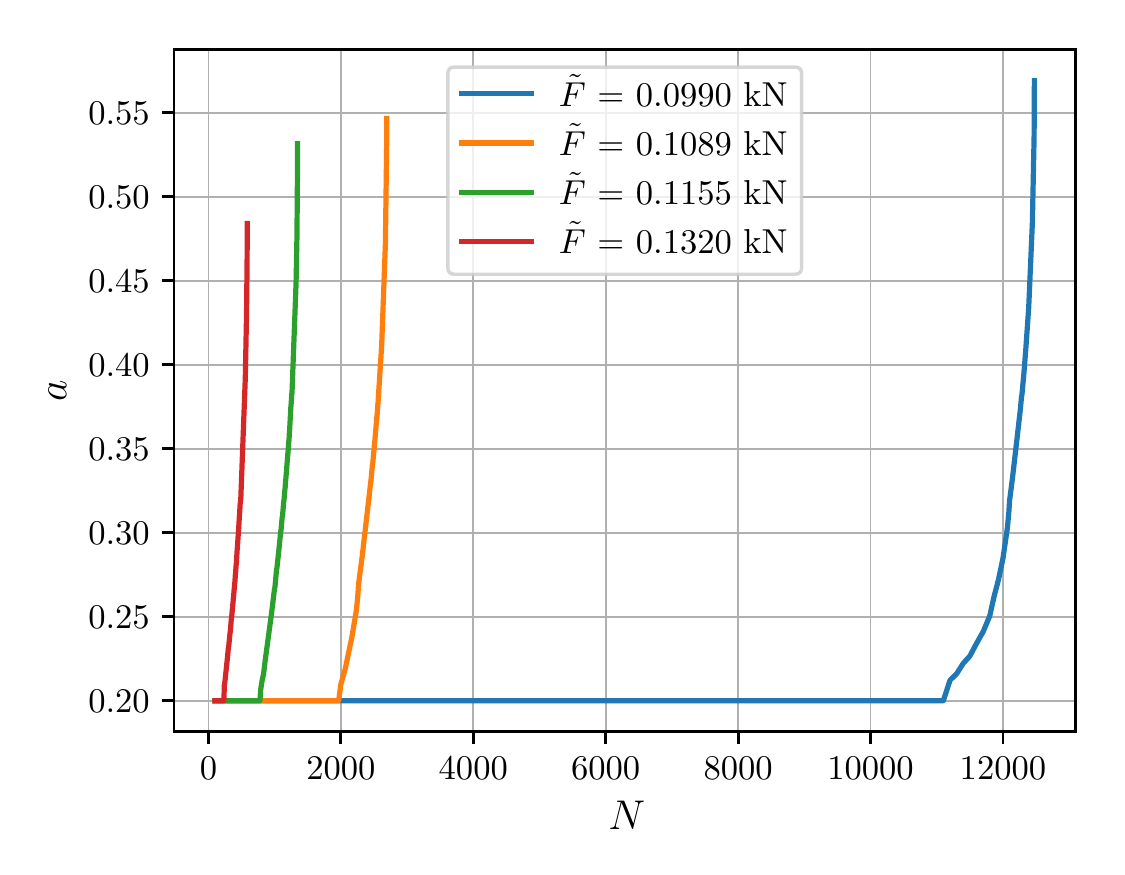} & a) &
		\includegraphics[width=0.5\linewidth]{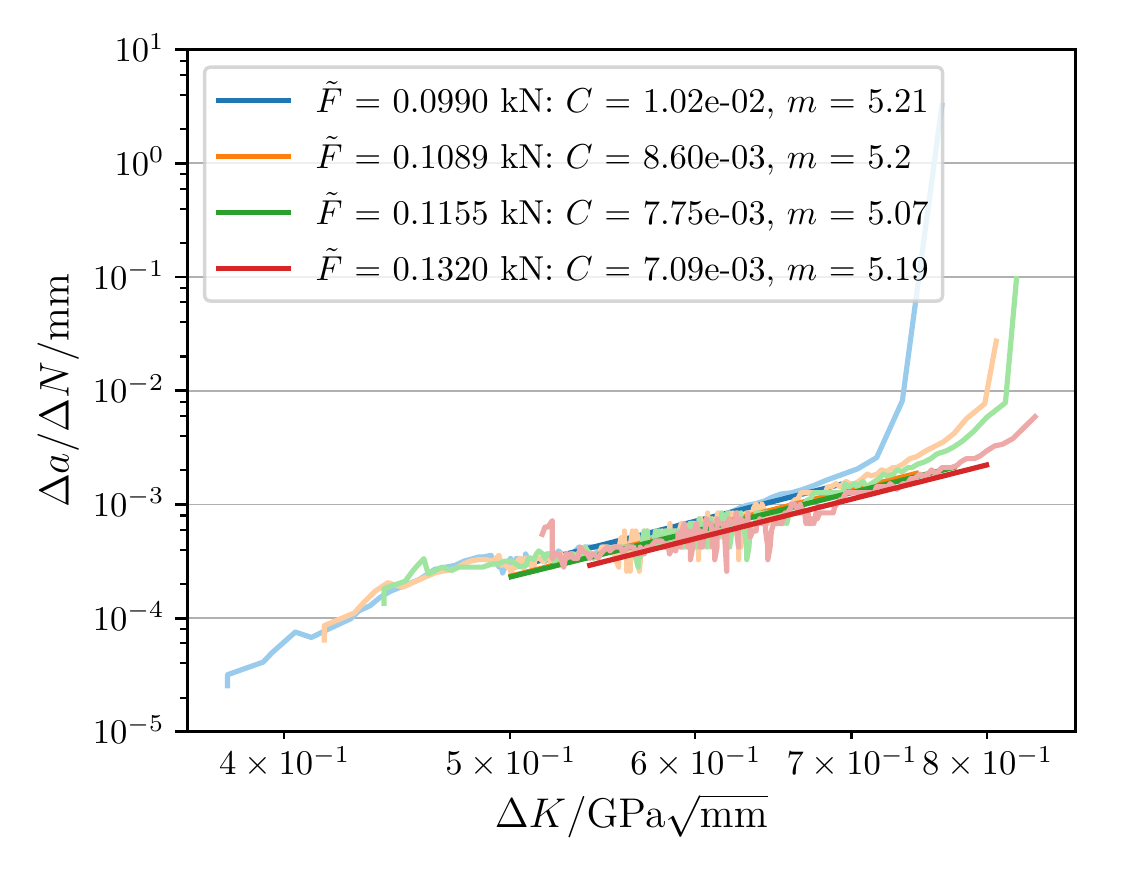} & b)\\
		\includegraphics[width=0.5\linewidth]{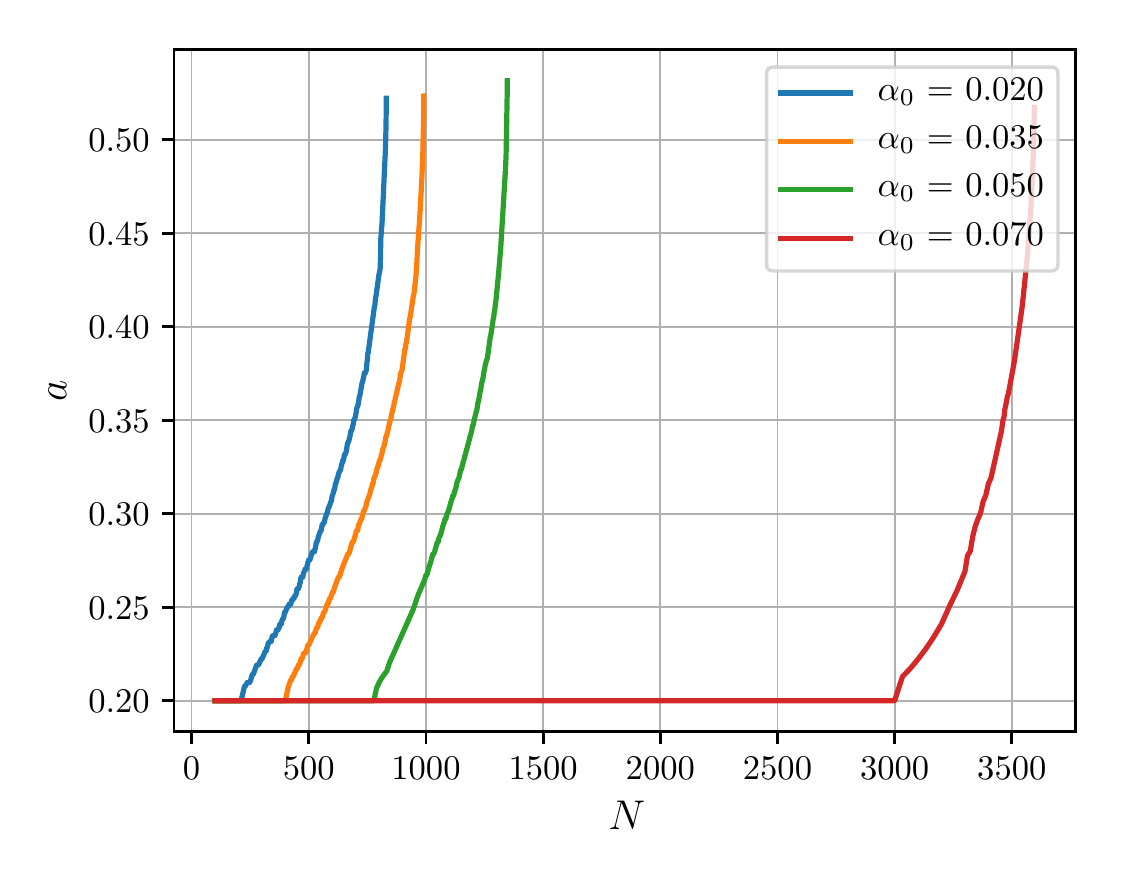} & c) &
		\includegraphics[width=0.5\linewidth]{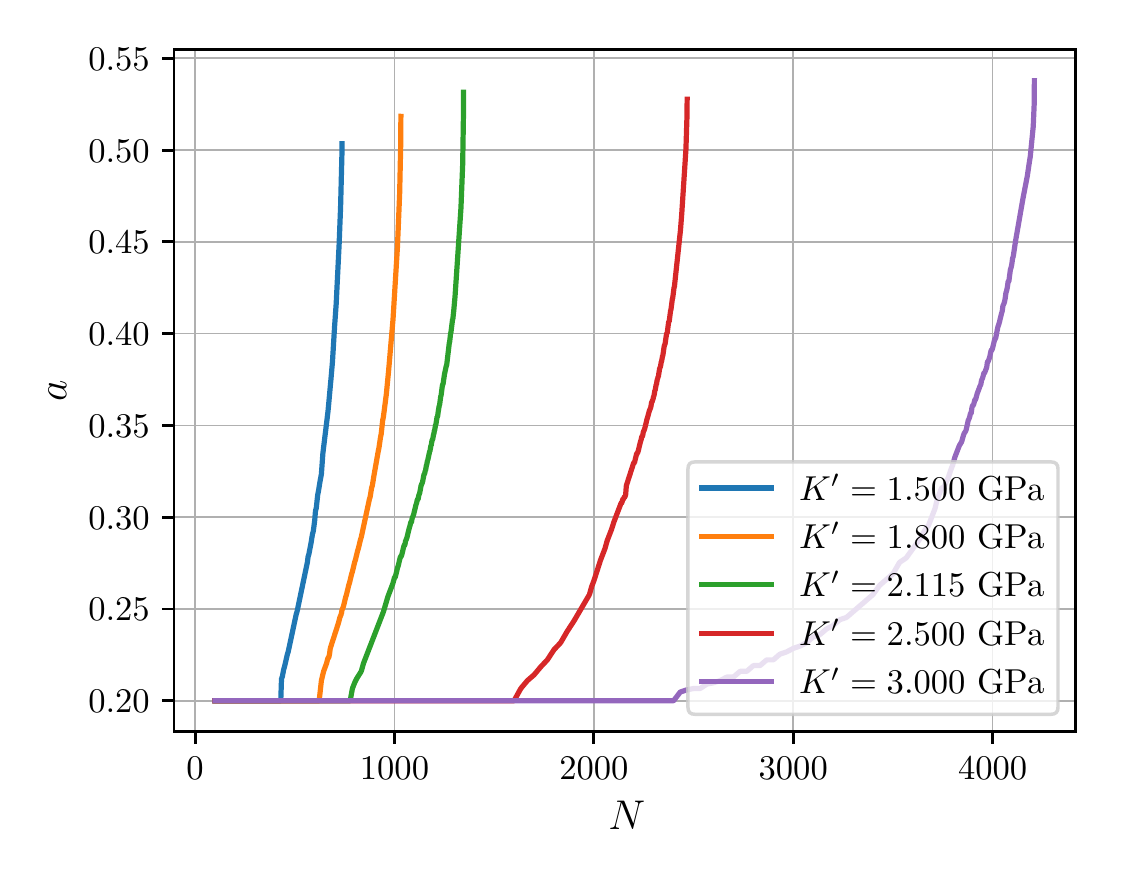} & d) \\
		\includegraphics[width=0.5\linewidth]{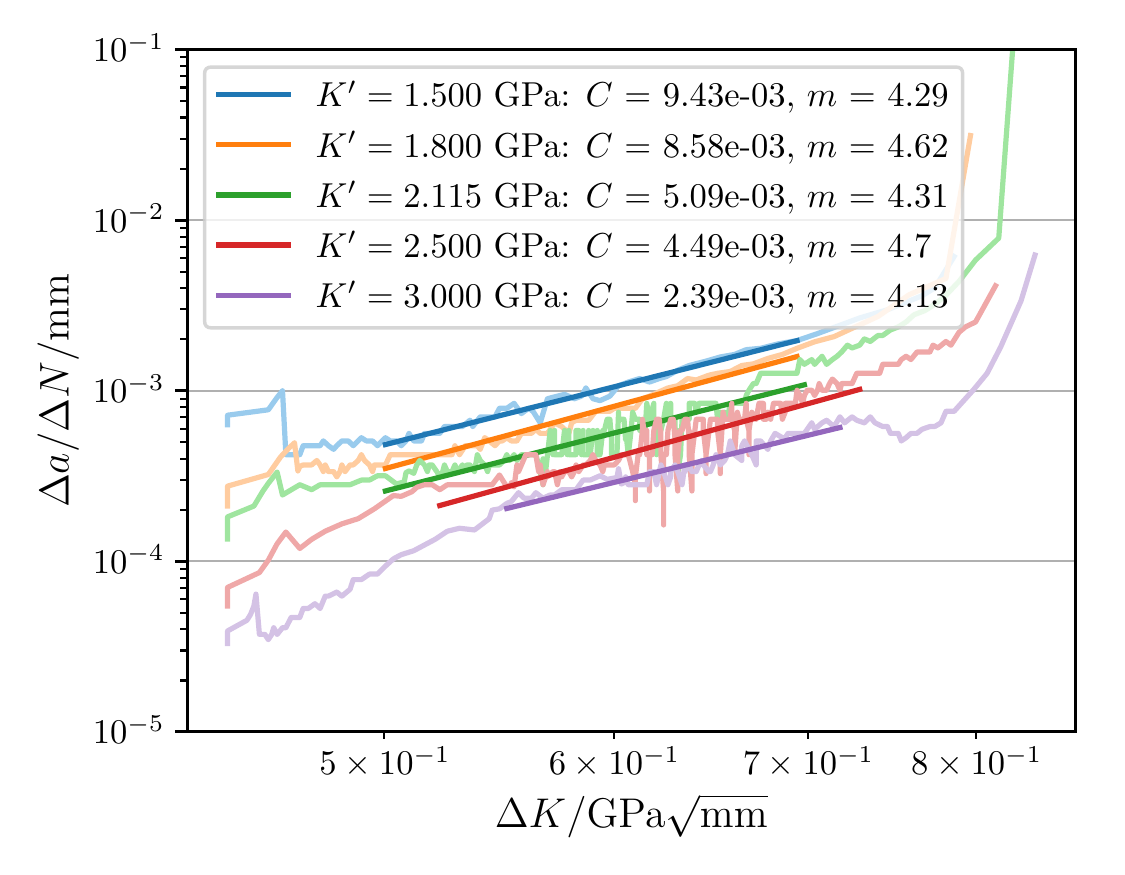} \hspace*{-2.6cm} \includegraphics[width=0.12\linewidth]{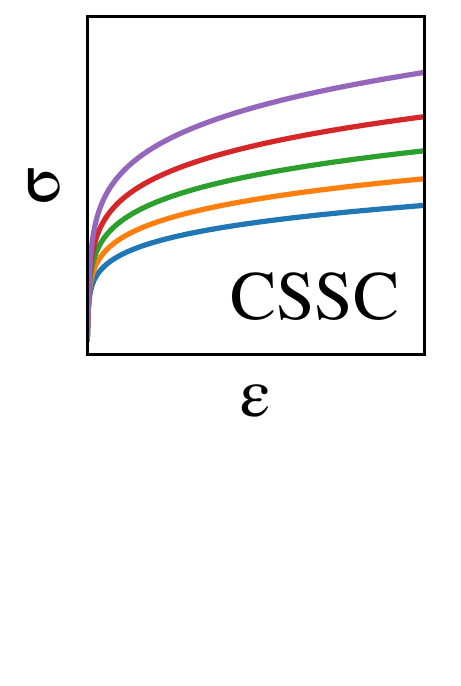}  & e) &
		\includegraphics[width=0.5\linewidth]{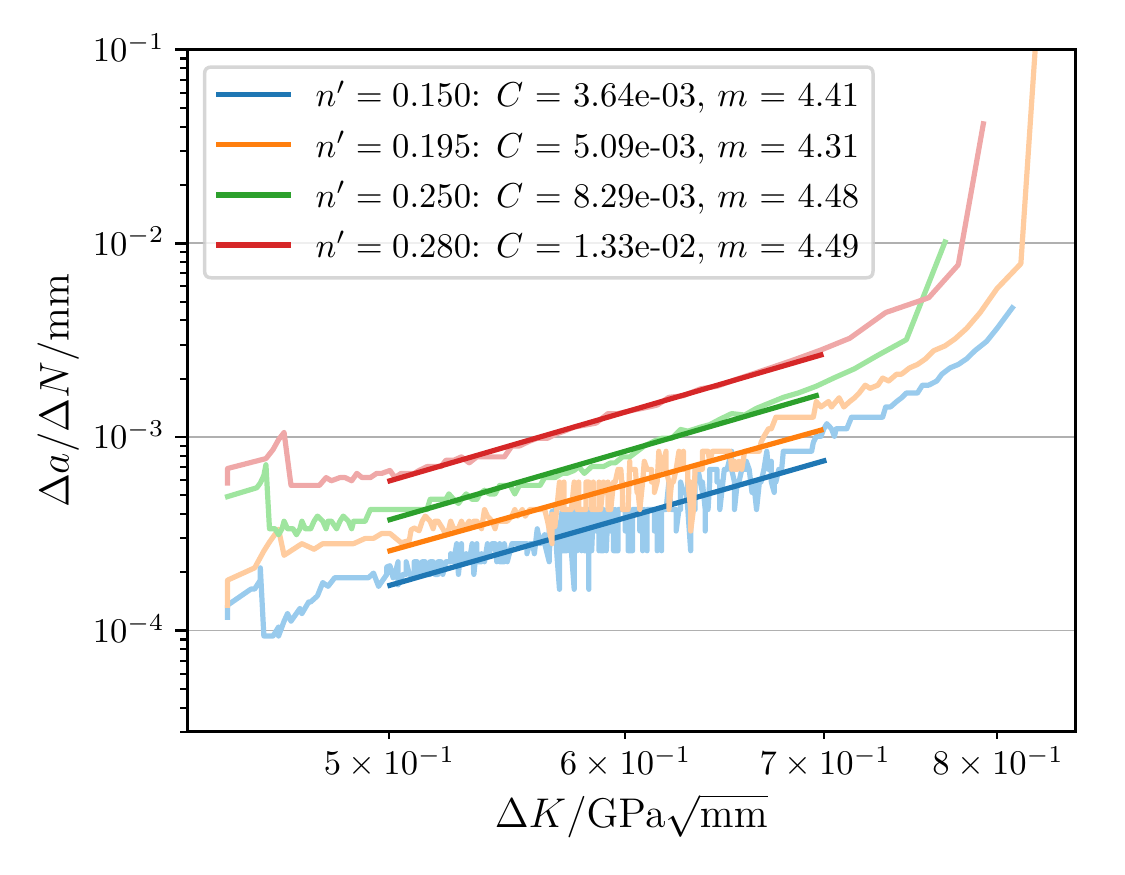} \hspace*{-2.6cm} \includegraphics[width=0.12\linewidth]{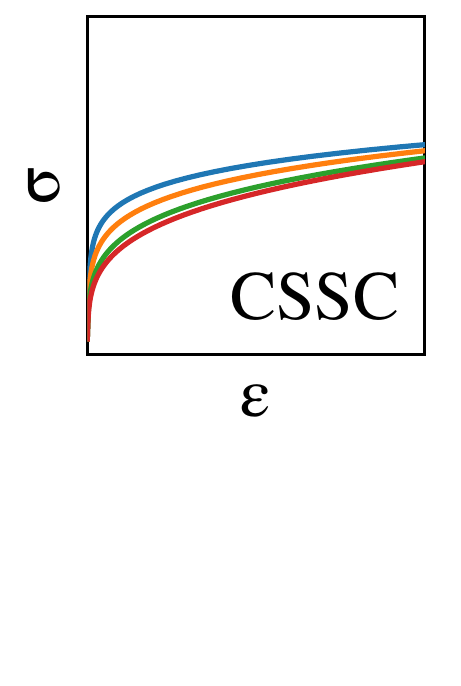} & f)
	\end{tabular}			 
	\caption{Cyclic compact tension test. Parameters: Load amplitude $\tilde{F}$, threshold of fatigue degradation function $\alpha_0$, parameters of cyclic stress-strain curve (CSSC) $K'$ and $n'$. \textbf{(a)}, \textbf{(c)}, \textbf{(d)} Crack length $a$ over number of load cycles $N$. \textbf{(b)}, \textbf{(e)}, \textbf{(f)} Paris plot: Crack propagation rate $\Delta a/\Delta N$ over amplitude of stress intensity factor at crack tip $\Delta K$. Fitted with the Paris parameters $C$ and $m$ within the Paris range. For $K'$ and $n'$, the CSSCs are schematically depicted.
		\label{fig:2D2}}
\end{figure}

The following parameter study is performed with the simplified model displayed in Fig.~\ref{fig:CT_setup}c). For the sake of simplicity, the free ends behind the borehole are not included in the model, instead the axes of load application are evened out. If not stated differently $\tilde{F}=0.1155$\,kN and $a_0=0\,W$ holds.

In Fig.~\ref{fig:2D1}c), the minimum element size is varied. Obviously, convergence is  reached for $\ell/h_\mathrm{min}=6.4$. Nevertheless, in all other simulations $\ell/h_\mathrm{min}=3.2$ is used, which already is fairly close to the converged value. Also in terms of load cycles per increment $\Delta N$ convergence should be studied. For this purpose, $\Delta N$ is held constant at first. Towards the end of the simulation $\Delta N$ is reduced which is marked by the dashed line. As shown in Fig.~\ref{fig:2D1}d), the simulations are in good alignment for varying $\Delta N $. 
Fig.~\ref{fig:2D1}e-f) show that the characteristic length $\ell$ has a significant influence on crack initiation -- the smaller $\ell$ is, the earlier a crack initiates -- but not on the Paris parameters. For a larger $\ell$, the "plastic zone" grows bigger.
\FloatBarrier

Fig.~\ref{fig:2D2}c) shows the influence of the threshold $\alpha_0$ of the fatigue degradation function. A higher $\alpha_0$ leads to delayed crack initiation. Since the fracture toughness is not reduced as much, the "plastic zone" has to grow bigger in order to initiate a crack.
The same can be shown for a varying load amplitude $\tilde{F}$, displayed in Fig.~\ref{fig:2D2}a-b). As expected, for higher loads the crack initiates earlier. Nevertheless, the fitted Paris parameters are roughly the same. This shows that the model can reproduce Paris behaviour with the material parameters $C$ and $m$ independent from loading.

However, according to Paris theory, material properties can change the Paris parameters. The cyclic hardening coefficient and exponent $K'$ and $n'$ of the CSSC according to the Ramberg-Osgood model (\ref{eq:RamOs}) are varied in Fig.~\ref{fig:2D2}d-f). The corresponding CSSCs are shown within the figures. Apparently, the Paris line is shifted upwards by increasing $n'$ and decreasing $K'$, respectively. Both lead to a lower CSSC. Hence, the revaluation for same stress and strain yields a higher plastic strain and therefore a wider stress-strain hysteresis. This is associated with more dissipation. Consequently the higher damage parameter leads to a stronger degradation of the fracture toughness. This causes higher crack propagation rates and higher Paris lines. Moreover, since already small stresses lead to large plastic strains, the degraded zone as well as the phase-field show a wider profile for increasing $n'$ and decreasing $K'$ and cracks initiate earlier.

\section{Conclusion}
\label{sec:Conc}

A combination of the phase-field method for brittle fracture with a fatigue life concept is introduced. It can model cyclic crack initiation and propagation as is demonstrated in 1D- and 2D-examples. The fatigue effects are considered by degrading the fracture toughness depending on a local lifetime variable. This variable is determined with the so-called local strain approach, considering plasticity as the cause of ductile fatigue fracture. Static fracture is included in the formulation as a special case. The model causes relatively small computational effort, since an elasto-plastic stress-revaluation is performed instead of using an elasto-plastic material model. Moreover, several load cycles can be simulated within one increment. It is shown that the model recovers Paris behaviour.

Further research will now concentrate on validation with experiments and the determination of the parameters of the fatigue degradation function. Moreover, the revaluation technique for elastic stresses has to be compared to more elaborate revaluation concepts and an elasto-plastic material model.

\section*{Acknowledgements}

This work was supported by the Deutsche Forschungsgemeinschaft in the Priority Program
2013 \textit{Targeted Use of Forming Induced Residual Stresses in Metal Components} (grant number KA 3309/7-1). We thank Jörg Brummund for his contributions.

\section*{Highlights} 

\begin{enumerate}
	\item A new phase-field model for fatigue crack initiation and propagation is proposed.
	\item It minimises computational effort due to a combination with a classic fatigue concept.
	\item A revaluation technique replaces an elasto-plastic material model.
	\item The model can reproduce Paris behaviour.
\end{enumerate}

\section*{References}

\bibliography{mybibfile}	

\end{document}